\documentclass[twocolumn,prb,citeautoscript,showpacs,amsmath,amstex,amssymb,mathfonts]{revtex4-1}
\pdfoutput=1
\usepackage{amsthm,color,amsfonts,graphicx,verbatim}
\usepackage{amsmath}
\usepackage{amssymb}
\usepackage{amsthm}
\usepackage{amsfonts}
\usepackage{listings}
\usepackage{enumerate}
\usepackage{latexsym}
\usepackage{bm}
\usepackage{graphicx}
\usepackage{color}
\usepackage{hyperref}
\usepackage{rotating}
\hypersetup{
    bookmarks=true,         
    unicode=false,          
    pdftoolbar=true,        
    pdfmenubar=true,        
    pdffitwindow=false,     
    pdfstartview={FitH},    
    pdftitle={Quantum quenches in the many-body localized phase},    
    pdfauthor={M. Serbyn, Z. Papic, D. Abanin},     
    pdfsubject={},   
    pdfcreator={},   
    pdfproducer={}, 
    pdfkeywords={quantum quench} {many-body localization} {many body dynamics}, 
    pdfnewwindow=true,      
    colorlinks=true,       
    linkcolor=black,          
    citecolor=blue,        
    filecolor=magenta,      
    urlcolor=blue           
}

\newcommand{\be}{\begin{equation}}
\newcommand{\ee}{\end{equation}}
\newcommand{\bea}{\begin{eqnarray}}
\newcommand{\eea}{\end{eqnarray}}

\newcommand{\la}{\langle}
\newcommand{\ra}{\rangle}

\renewcommand{\phi}{\varphi}
\renewcommand{\epsilon}{\varepsilon}

\newcommand{\corr}[1]{\langle{ #1}\rangle}
\begin{document}
\title{Quantum quenches in the many-body localized phase}
\author{Maksym Serbyn$^1$, Z. Papi\'c$^{2,3}$, and D. A. Abanin$^{2,3}$}
\affiliation{$^1$ Department of Physics, Massachusetts Institute of Technology, Cambridge, MA 02138, USA}
\affiliation{$^2$ Perimeter Institute for Theoretical Physics, Waterloo, ON N2L 2Y5, Canada}
\affiliation{$^3$ Institute for Quantum Computing, Waterloo, ON N2L 3G1, Canada}

\date{\today}
\begin{abstract}

Many-body localized (MBL) systems are characterized by the absence of transport and thermalization, 
and therefore cannot be described by conventional statistical mechanics. In this paper, using analytic
arguments and numerical simulations, we study 
the behaviour of local observables in an isolated MBL system following a quantum quench. 
For the case of a global quench, we find that the local observables reach stationary, 
highly non-thermal values at long times as a result of slow dephasing characteristic of the MBL phase. 
These stationary values retain the local memory of the initial state due to the existence of local integrals
of motion in the MBL phase. The temporal fluctuations around stationary values exhibit universal power-law decay in time, with an exponent set by the localization length and the diagonal entropy of the initial state. Such a power-law decay holds for any local observable and is related to the logarithmic in time growth of entanglement in the MBL phase. This behaviour 
distinguishes the MBL phase from both the Anderson insulator (where no stationary state is reached), and from the ergodic phase (where relaxation is expected to be exponential). For the case of a local quench, we also find a power-law approach of local observables to their stationary values when the system is prepared in a mixed state.
Quench protocols considered in this paper can be naturally implemented in systems of ultra cold atoms in disordered optical lattices, and the behaviour of local observables provides a direct experimental signature of many-body localization. 

\end{abstract}

\maketitle

\section{Introduction}

Over the past decade, there has been significant progress in understanding dynamics, and in particular, the mechanisms of thermalization and its breakdown in closed quantum many-body systems~\cite{PolkovnikovColdAtoms}. The renewed interest in the emergence and limitations of statistical mechanics has largely been inspired by the revolution in experimental techniques, which lead to the realization of isolated, tunable quantum many-body systems of cold atoms~\cite{BlochColdAtoms}, trapped ions~\cite{Blatt12}, and superconducting qubits~\cite{Devoret13}. Such systems allow one to create non-equilibrium many-body states and characterize their unitary evolution via measurements of physical observables.  

At this point, there is numerical~\cite{RigolNature} and experimental (see Ref.~[\onlinecite{PolkovnikovColdAtoms}] for a review) evidence that certain closed quantum systems (which we refer to as ergodic) do thermalize as a result of unitary evolution, despite always being in a pure state: local observables reach stationary values which are determined only by the global characteristics of the initial state (e.g., energy and particle number). It is believed that the mechanism underlying thermalization is the ``eigenstate thermalization hypothesis" (ETH)~\cite{DeutschETH, SrednickiETH}, which states that individual many-body eigenstates of ergodic systems are locally thermal, with the observables described by an appropriate Gibbs ensemble.

However, not all systems thermalize; much work addressed the dynamics in Bethe Ansatz integrable systems, which are characterized by an infinite number of integrals of motion~\cite{Sutherland}. In a pioneering experiment, Kinoshita \emph{et al.}~\cite{Kinoshita06} demonstrated the lack of complete thermalization in an integrable system of 1D bosons. Theoretical works~\cite{Rigol07,GGE2,GGE3} indicate that in integrable systems stationary values of local observables are given by so called ``generalized Gibbs ensemble", which accounts for the additional conservation laws. 

Recently, it has become evident that there exists another general class of many-body systems which break ergodicity and fail to thermalize -- the many-body localized (MBL) systems~\cite{Basko06,Mirlin05,OganesyanHuse,PalHuse}.  Many-body localization is driven by disorder, and loosely can be viewed as Anderson localization~\cite{Anderson58} in the many-body Hilbert space. Similar to the Anderson insulator of non-interacting particles, when decoupled from a thermal bath the MBL systems do not conduct heat or particle number currents, and therefore cannot fully thermalize. 

It was recently shown that the lack of thermalization in the MBL phase can be linked to the existence of an extensive number of emergent integrals of motion, which strongly restrict quantum dynamics.~\cite{Serbyn13-1,HuseOganesyan13} These integrals of motion are \emph{local}, in the sense that they act non-trivially only on a small number of physical degrees of 
freedom.~\footnote{The notion of locality used here is different from the one used in conventional 1D integrable systems. Here, locality implies that an operator acts non-trivially on $\sim O(1)$  degrees of freedom. In this sense, a spin operator on a given site is local, whereas the Hamiltonian is not. Integrals of motion in the context of integrable systems are called local when they can be expressed as a sum of a local operators, and both examples above satisfy this definition.} 
In systems where all the many-body states are localized, local integrals of motion form a complete set, as their number is equal to the number of physical degrees of freedom. For example, in an MBL system of $N$ interacting spins, it is possible to define $N$ local conserved ``effective" spins. A key property of MBL systems compared to the Bethe Ansatz integrable systems is their robustness: if the Hamiltonian of an MBL system is slightly perturbed, it remains in the MBL phase, and a new set of local integrals of motion can be defined. In contrast, even small generic perturbations can break the integrability by Bethe Ansatz. Thus, many-body localization gives rise to a new class of non-ergodic phases of matter. 

The existence of an extensive set of local integrals of motion leads to novel dynamical properties of the MBL phase, which distinguish it from ergodic and Bethe Ansatz integrable systems. The main goal of this paper is to study the experimentally measurable signatures of the dynamics in the MBL phase. To that end, we study an MBL system subject to an instantaneous quantum quench -- a standard setup used to characterize quantum many-body dynamics~(see e.g. Ref.~[\onlinecite{PolkovnikovColdAtoms}]). We focus on the behaviour of local observables. We argue that, as a result of residual interactions in the MBL phase which give rise to dephasing~\cite{Serbyn13-1,Serbyn13-2,Altman13,HuseOganesyan13}, the system reaches a highly non-thermal stationary state at long times. {\it Any} local observable reaches a stationary value at long times, which retains the memory of the initial state. 

Temporal fluctuations of local observables around their long-time value generally decay universally, according to a power-law with an exponent set by the localization length and the properties of the initial state (the density of its second diagonal Renyi entropy). Power-law relaxation is a characteristic feature of an MBL system,  which distinguishes it from both the Anderson insulator (where relaxation is absent), and from the ergodic phase (where relaxation is expected to be exponential).  

The power-law temporal fluctuations of local observables following a quantum quench provide a direct experimental signature of the dynamics in the MBL phase, which is limited to an exponentially slow dephasing between remote degrees of freedom. Previous works~\cite{Serbyn13-1, Serbyn13-2,HuseOganesyan13,Altman13} showed that such dephasing underlies the logarithmic-in-time growth of entanglement for initial product states~\cite{Znidaric08, Moore12, Serbyn13-2}. This should be contrasted with ballistic spreading of entanglement and quantum correlations in Bethe-ansatz-integrable systems (according to standard Lieb-Robinson bounds~\cite{LiebRobinson,Bloch12,KimHuse13}). We note that an alternative way of probing MBL in a  spin-echo-type experiment was recently suggested in Ref.~[\onlinecite{Serbyn14}]. Also, in Ref.~[\onlinecite{Vasseur14}], it was shown that the dephasing dynamics in the MBL phase leads to characteristic revivals of local observables. 
  
The paper is organized as follows. In Sec.~\ref{Sec:model} we review the effective description of the MBL phase based on the picture of local integrals of motion~\cite{Serbyn13-2, HuseOganesyan13}. In Sec.~\ref{Sec:modelXXZ} we introduce the physical model that will be the subject of our numerical studies. Next, in Sec.~\ref{Sec:quench}, we discuss the physics of a global quench within the effective model of the MBL phase. In particular, we analyze the behaviour of single- and multi-spin operators following a quench in one-dimensional MBL systems, and comment on higher dimensions and different types of initial states. In Sec.~\ref{sec:general} we study the behaviour of physical observables following a global quench and present results of numerical simulations. Sec.~\ref{Sec:quench2} is devoted to local quenches, while Sec.~\ref{Sec:concl} contains the summary of our results and concluding remarks.

 \section{Effective model of the many-body localized phase \label{Sec:model}} 
 
Throughout the paper, we consider MBL systems on a lattice that may naturally be realized in disordered optical lattices~\cite{DeMarco10,Inguscio13,DeMarco13}, systems of trapped ions, and NV-centers in diamond~\cite{Lukin06, Wrachtrup08}. We will assume that all states in the many-body spectrum of our system are localized, that is, many-body localization persists up to an infinite temperature~\cite{OganesyanHuse}. Numerical studies~\cite{PalHuse, Serbyn13-2, BauerNayak, Kjall14} have provided support for the existence of such an infinite-temperature MBL phase in certain disordered spin models. One such model will be described in Sec.~\ref{Sec:modelXXZ} below.
 
We will analyze the dynamical properties of the MBL phase using an effective model introduced in Refs.~[\onlinecite{Serbyn13-1},\onlinecite{HuseOganesyan13}]. This model is based on the hypothesis that the MBL phase is characterized by a complete set of local integrals of motion, with exponentially decaying interactions between them. The validity of this picture was recently proven rigorously for certain spin chains~\cite{Imbrie14}, and local integrals of motion have been explicitly constructed numerically~\cite{Chandran14} and perturbatively~\cite{Ros14}. 
 
The choice and representation of the local conserved quantities is not unique. For our purposes, it is convenient to use the spin-$1/2$ representation. In an interacting chain of $L$ spins $1/2$, with Hilbert space dimension $2^L$, we can choose $L$ operators $\tau_i^z$ which can be viewed as $z$-projections of some "effective" spins ("l-bits" in terminology of Ref.~[\onlinecite{HuseOganesyan13}]), and satisfy the following properties: (i) Each $\tau_i^z$ commutes with the Hamiltonian $\hat H$, and other $\tau^z$ operators: $[\tau_i^z, \hat H]=0$, $[\tau_i^z, \tau_j^z]=0$; (ii) Each $\tau_i^z$ is local: it strongly affects only the degrees of freedom in some finite region of space (support), and its effect on remote regions decays exponentially with the distance from its support. We will choose labels $i$ in such a way that they reflect the spatial structure of the local integrals of motion, and $|i-j|$ is proportional to the distance between supports of $\tau_i^z,\tau_j^z$. In an MBL spin system, $\tau_i^z$ can be viewed as the physical spin operator $\sigma_i^z$ dressed with a quasi-local unitary transformation. The role of this unitary transformation is to ``unwind" the eigenstates of $\hat{H}$ into product states, and its existence is closely related to the fact that nearly all MBL eigenstates obey a boundary law for entanglement entropy~\cite{Serbyn13-1, BauerNayak}.

In terms of $\tau$-operators, the effective MBL Hamiltonian takes the following universal form:
\be\label{eq:Hamiltonian}
\hat H=\sum_i h_i \tau_i^z+\sum_{i,j} J_{ij} \tau_i^z \tau_j^z+\sum_{i,j,k} J_{ijk} \tau_i^z \tau_j^z \tau_k^z+...
\ee
Here $h_i$ describes the random field acting on the effective spin $i$. A salient feature of this Hamiltonian is the absence of any hopping terms, as it only depends on $\tau^z$-operators (and not on $\tau^x,\tau^y$). The spin-spin interaction terms arise when interactions are present in the original Hamiltonian, and would be absent in a non-interacting Anderson insulator. These terms are expected to be random and to decay exponentially with distance $r_{ij}=|i-j|$, with some characteristic  length scale~$\xi_1$ 
\be\label{eq:interactions}
J_{ij}\propto J_0 e^{-\frac{r_{ij}}{\xi_1}}, 
\quad 
J_{ijk}\propto  J_0 e^{-\frac{{\rm max}(r_{ij},r_{jk},r_{ik})}{\xi_1}},\ldots
\ee 
where $J_0$ sets the interaction scale.  

In the $\tau$-basis, eigenstates of the Hamiltonian (\ref{eq:Hamiltonian}) are simply product states with $\tau_{i}^z=\pm 1$. In the physical basis, eigenstates are entangled, but only slightly, and almost all of them obey a boundary law for entropy~\cite{Serbyn13-1, BauerNayak}. This unusual property, typically found in ground states of gapped systems, here occurs in any excited state. In contrast, excited eigenstates of ergodic systems are far more entangled: their entanglement entropy scales according to a volume law. 

Despite the simple form of the Hamiltonian in the $\tau$-basis, the MBL phase exhibits non-trivial dynamics when the system is initially prepared in a superposition of eigenstates. Each effective spin experiences an effective magnetic field in the $z$-direction that depends on the state of other spins in a complicated way. Therefore, if the system is initially prepared in a state where each effective spin is in a superposition state of $\tau_i^z=\pm 1$, over time different spins will get entangled. This entanglement is caused by the dephasing of the off-diagonal elements of the reduced density matrix in the $\tau$-basis~\cite{Serbyn13-2,HuseOganesyan13,Serbyn13-1}. 

The magnetic field experienced by the effective spin $k$ can be represented as follows: 
\be\label{eq:Heff}
H_{k}(\{ \tau' \})=h_k+H_k^1(\{\tau' \})+H_{k}^2(\{\tau' \})+\ldots,
\ee
where $\{\tau'\}$ refers to the configuration of all spins other than $k$th. $H_{k}^l$ denotes the magnetic field arising from interactions with the spins which are within the distance $|i-k|\leq l$ from spin $k$, e.g., the first two terms are explicitly given by
\be\label{eq:H1}
H_k^1=J_{k,k+1}\tau_{k+1}^z+J_{k,k-1}\tau_{k-1}^z + J_{k,k-1,k+1} \tau_{k-1}^z\tau_{k+1}^z, 
\ee
and
\begin{eqnarray}\label{eq:H2}
\nonumber H_k^2 &=& J_{k,k+2}\tau_{k+2}^z+J_{k,k-2}\tau_{k-2}^z \\
\nonumber  &+& \sum_{|\sigma| \leq 2}^{'} \sum_{2\geq \sigma' > \sigma}^{'} J_{k,k+\sigma,k+\sigma'} \tau_{k+\sigma}^z \tau_{k+\sigma'}^z \\
\nonumber &+& \sum_{|\sigma| \leq 2}^{'} \sum_{2\geq \sigma' > \sigma}^{'} \sum_{2 \geq \sigma'' > \sigma' }^{'} J_{k,k+\sigma,k+\sigma',k+\sigma''} \tau_{k+\sigma}^z \tau_{k+\sigma'}^z \tau_{k+\sigma''}^z \\
&+& J_{k-2,k-1,k,k+1,k+2} \tau_{k-2}^z \tau_{k-1}^z \tau_{k+1}^z \tau_{k+2}^z, 
\end{eqnarray}
where the prime in $\sum'$ indicates that the summations exclude zero ($\sigma\neq 0$, etc.).

It is believed that the effective fields $H_{k}^l$ decay exponentially with distance $l$: 
\be\label{eq:decay}
H_{k}^l \sim J_0 e^{-l/\xi}, 
\ee
with a characteristic scale $\xi$ that generally differs from the length scale $\xi_1$ that controls the decay of interactions in Eq.~(\ref{eq:interactions}).  In principle, the two length scales $\xi,\xi_1$ can be related using the form of $H_{k}^l$; however, for the sake of simplicity we will treat them as phenomenological parameters of the theory.

 \section{Microscopic model \label{Sec:modelXXZ}}
 
Now we briefly introduce a microscopic model of the MBL phase, which will be studied numerically in Sec.~\ref{Sec:sim}. Namely, we consider a 1D XXZ spin chain in a random magnetic field, which is believed to exhibit the MBL phase at sufficiently strong disorder~\cite{PalHuse}. The Hamiltonian for 1/2-spins is given by:
 \begin{equation} \label{Eq:XXZ}
\hat H_\text{XXZ} = J_\perp \sum_{\langle ij \rangle} (s_i^x s_j^x + s_i^y s_j^y) 
 + J_z \sum_{\langle ij \rangle} s_i^z s_j^z +  \sum_i w_i  s_i^z,
\end{equation}
where the disorder enters via random field $w_i$, which we take to be uniformly distributed in the interval $ [-W;W]$. Interaction and hopping terms extend over the nearest neighbor spins only. For open boundary conditions and $J_z=0$, the model (\ref{Eq:XXZ}) is equivalent to free fermions moving in a disorder potential, via Jordan-Wigner transformation. In this limit, the system is in the Anderson-localized phase for any $W > 0$. When $J_z \neq 0$, the system is believed to exhibit both MBL and delocalized phases as a function of $W/J_\perp$~\cite{PalHuse}. In particular, for $J_\perp=J_z=1$, the phase boundary was identified to be located at $W^* \approx 3.5 \pm 1.0$. 

 We emphasize that the Hamiltonian (\ref{Eq:XXZ}) in the MBL phase can be brought into the form~(\ref{eq:Hamiltonian})  by an appropriate unitary transformation. However, the effective spins operators have a complicated and non-universal, albeit local,  relation to the physical spin operators. In what follows, we will study local observables for both effective and physical spin operators analytically, and find that their dynamics is similar. Later, we use model (\ref{Eq:XXZ}) to compute behavior of physical spin observables numerically, providing additional test of analytic results.

\section{Global quench within the effective model \label{Sec:quench}}
 
Now we proceed to discuss global quantum quenches in the MBL phase. We will assume that the system is initially prepared in a product state in the physical basis, or more generally, in a weakly entangled state. Such states, in general, are superpositions of different eigenstates. At $t>0$ the system is evolved with the MBL Hamiltonian~(\ref{eq:Hamiltonian}). Later on, we will also consider the case when the system was initially in the ground state of some other Hamiltonian, and then its Hamiltonian was abruptly switched to $\hat H$ at $t=0$. Throughout the analysis, we assume that the system is isolated and decoupled from an external bath, or that the time scale of interaction with the bath is longer than the observation time.

In this Section, we start by considering a particularly simple initial state in which {\it effective} (rather than physical) spins are prepared in a product state. We assume that each effective spin initially points in some direction on the Bloch sphere. In practice, such a state is hard to prepare, as the relation of effective spins to physical degrees of freedom is disorder-realization-dependent and {\it a priori} not known. However, this example captures most of the key features of the quench dynamics in the MBL phase, and has the advantage of being analytically tractable. Below we start by considering expectation values of a single spin operator (Sec.~\ref{singlespin}), and later generalize to multi-spin operators (Sec.~\ref{multispin}).

\subsection{Single-spin observables \label{singlespin}}

The initial state is given by 
\be\label{eq:initial}
|\Psi (t=0)\ra=\otimes_{i=1}^L (A_{i\uparrow}|\!\uparrow\ra _i+A_{i\downarrow}|\!\downarrow\ra _i), 
\ee 
where $|\!\!\uparrow\!\!(\downarrow)\ra _i$ denotes $\tau_i^z=\pm 1$ states, and $A_{i\uparrow},A_{i\downarrow}$ are complex numbers satisfying a normalization condition $|A_{i\uparrow}|^2+|A_{i\downarrow}|^2=1$. Under unitary evolution with the Hamiltonian~(\ref{eq:Hamiltonian}), different eigenstates entering the wave function (\ref{eq:initial}) acquire different phases. The wave function at time $t$ is given by: 
\be\label{eq:timet}
|\Psi(t)\ra=\sum_{\{\tau \}} \left( \prod_{i=1}^L  A_{i\tau_i} \right) e^{-iE_{\{ \tau \}}t} |\{\tau \}\ra, 
\ee
where $|\{ \tau\}\ra$ is an eigenstate of the Hamiltonian (\ref{eq:Hamiltonian}), with a given configuration of effective spins, e.g. $\{\tau\}=\uparrow\downarrow\downarrow\uparrow\ldots\uparrow$, and $E_{\{ \tau \}}$ is the energy of such a state. 

We consider single-spin observables for spin $k$, described by the operators $\tau_k^\alpha(t)$, $\alpha=x,y,z$. The diagonal elements of the reduced density matrix for spin $k$ are time-independent because $\tau_k^z$ is an integral of motion: 
\be\label{eq:reduced1}
\rho_{\uparrow\uparrow}(t)=|A_{k\uparrow}|^2, \,\, \rho_{\downarrow\downarrow}(t)=|A_{k\downarrow}|^2, 
\ee
while the off-diagonal element of the density matrix reads: 
\be\label{eq:reduced2}
\rho_{\uparrow\downarrow}(t)=\rho^*_{\downarrow\uparrow}(t)=A_{k\uparrow}A_{k\downarrow}^* \sum_{\{ \tau'\}} P_{\{ \tau' \}}  e^{i\left(E_{\uparrow,\{\tau' \}}-E_{\downarrow,\{\tau' \}}\right)t}, 
\ee
where $\{\tau' \}$ refers to all configurations of $L-1$ spins with the $k$-th spin excluded. $E_{\uparrow(\downarrow),\{ \tau'\}}$ is the energy of a state in which $\tau_k^z=\uparrow\!\!(\downarrow)$, and the remaining spins are in a state $\{ \tau' \}$.  $P_{\{\tau' \}}$ is the probability of finding such a state, which is conserved during the unitary evolution. For the initial state (\ref{eq:initial}), this probability is given by $P_{\{ \tau' \}}=\prod_{i\neq k} |A_{i \tau_i}|^2$. 

Using Hamiltonian (\ref{eq:Hamiltonian}), we rewrite Eq.~(\ref{eq:reduced2}) as
\be\label{eq:reduced2-1}
\rho_{\uparrow\downarrow}(t)=A_{k\uparrow}A_{k\downarrow}^* \sum_{\{ \tau'\}}P_{\{ \tau' \}} e^{2iH_k(\{ \tau'\})t}, 
\ee
where we used $H_k(\{ \tau' \})$ -- the effective magnetic field experienced by the spin $k$ when the remaining spins are in a state $|\{\tau' \}\ra$ -- defined in Eq.~(\ref{eq:Heff}). 

In the absence of interactions, the effective magnetic field experienced by the $k$th spin does not depend on the state of other spins, therefore $H_{k}=h_k$ and $\rho_{\uparrow\downarrow}(t)=A_{k\uparrow}A_{k\downarrow}^* e^{2i h_k t} $. This describes precession of the spin $k$, without any dephasing. In this case, single-spin observables keep oscillating in time, and no steady state is reached, which is the dynamical signature of the Anderson insulator.

In contrast, in the MBL phase the presence of interactions makes $H_k(\{ \tau' \})$ dependent on the configuration $\{ \tau' \}$. Different configurations of surrounding spins correspond to a different magnetic field experienced by the $k$th spin,  leading to the dephasing and suppression of the off-diagonal element $\rho_{\uparrow\downarrow}$ of the reduced density matrix. To estimate $\rho_{\uparrow\downarrow}(t)$, we use the hierarchical structure of the effective magnetic field, which follows from Eq.~(\ref{eq:Heff}). In the limit of short localization length, each successive term in the sum in Eq.~(\ref{eq:Heff}) is typically much smaller than the previous one: $|H_k^l|\gg |H_k^{l+1}| \gg\ldots $. This leads to the separation of time scales: at time $t$, such that
\begin{eqnarray}\label{timetoentangle}
1/H_{k}^{l} \lesssim t  \lesssim  1/H_k^{l+1},
\end{eqnarray}
the magnetic field is effectively independent of the state of spins for which $r_{ik}>l$, but configurations which differ in one or more spins for which $r_{jk}\leq l$ pick up a phase difference much greater than $2\pi$. Physically, this means that spin $k$ gets entangled with spins $k-l,\ldots,k-1,k+1,\ldots k+l$, i.e. those within a distance $l$. This implies that at time $t$ the off-diagonal element of the reduced density matrix, Eq.~(\ref{eq:reduced2}), consists of $N(t)=2^{2l}$ terms with random phases, i.e., the number of configurations of spins for which $r_{ik}\leq l$. The value of this sum depends on the probabilities $P_{\{ \tau'\}}$, which are determined by the amplitudes~$A_{i\tau_i}$. 
 
Assuming that all $P_{\{ \tau' \}}$ are approximately equal (e.g., all probabilities $P_{\{ \tau' \}}$ are equal when $|A_{i\tau_i}|=\frac{1}{\sqrt{2}}$, that is, when all the spins are initially polarized in $xy$-plane), the magnitude of $\rho_{\uparrow\downarrow}(t)$ can be estimated as follows:
\be\label{eq:estimate}
|\rho_{\uparrow\downarrow}(t)|\sim  \frac{|A_{k\uparrow}A_{k\downarrow}^*|}{\sqrt{N(t)}}. 
\ee
Using $N(t)=2^{2l}$ and the relation $l\sim \xi \log (J_0 t)$, which follows from Eqs.~(\ref{eq:decay}) and (\ref{timetoentangle}), we obtain: 
\be\label{eq:estimate2}
|\rho_{\uparrow\downarrow}(t)|\sim  \frac{|A_{k\uparrow}A_{k\downarrow}^*|}{(tJ_0)^{a}}, 
\quad 
 a=\xi \ln 2.
\ee
Thus, the off-diagonal element of the reduced density matrix decays as a power-law in time, with an exponent set by $\xi$. Note that the factor $\ln 2$ multiplying $\xi$ is non-universal. It depends on the details of the initial state, and equals $\ln 2$ only for the states for which probabilities of all configurations are (approximately) equal. 

For general initial states, the behaviour of $\rho_{\uparrow\downarrow}(t)$ is determined by the distribution of the coefficients $A_{i\tau_i}$ and the corresponding density of the diagonal Renyi entropy, as we now show. To establish this connection, note that the time-averaged value of the squared off-diagonal element of the reduced density matrix at time $t$ is given by 
\begin{equation}
\la \rho_{\uparrow\downarrow}^2(t)\ra =|A_{k\uparrow}A_{k\downarrow}^*|^2 \sum_{\{ \tau '_l  \}, |i-k|\leq l} P_{\{ \tau '_l \}}^2,   
\end{equation}
where the sum is taken over configurations of spins $\{ \tau'_l\}$ such that $1\leq|i-k|\leq l$. 
In order to obtain the above equation, we assumed that the phases generated due to interactions with spins $|i-k|\leq l$ are all random. The expression on the r.-h.s. of the above equation can be related to the second diagonal Renyi entropy $S_2$ of the region of size $2l$, obtained by expanding the initial state in the basis of eigenstates, as follows:
\begin{equation}
\sum_{\{\tau'_l \}} P_{\{ \tau '_l\}}^2=\exp(-S_2(2l)), 
\end{equation}
where $\{ \tau'_l \}$ denotes the configurations of spins situated within distance $l$ from spin $k$. Noting that $S(2l)$ is extensive and therefore $S(2l)=C\cdot 2l$ with a coefficient dependent on the initial state, and using $l\sim \xi \log (J_0 t)$, we obtain  
\be\label{eq:decay_general}
|\rho_{\uparrow\downarrow}(t)| \sim \frac{|A_{k\uparrow}A_{k\downarrow}^*|} {(tJ_0)^b}, \,\, b=\xi C. 
\ee
For the case of the maximum possible Renyi entropy, $C=\ln 2$, and this expression reduces to Eq.~(\ref{eq:estimate2}), therefore $b$ is bounded from above by $a$, $b\leq a = \xi\ln2$.

The results (\ref{eq:reduced1}) and (\ref{eq:decay_general}) for the components of the density matrix allow us to understand the time evolution of the single-spin observables: $\tau_k^z$ is conserved, and its expectation value does not change, while $\tau_k^x, \tau_k^y$ show a power-law decay in time to zero: 
\begin{eqnarray}\label{eq:tauz}
\la \tau_k^z(t)\ra&=&\la  \tau_k^z(0) \ra,
\\ \label{eq:taux}
|\la \tau_k^{x,y}(t)\ra| &\propto& \frac{1}{t^b}, 
\quad t\gg 1/J_0. 
\end{eqnarray}
Observables $\tau_k^{x,y}$ show fast non-universal oscillations, while the amplitude of the oscillations decays as a power-law in time. Below we will argue that this behaviour holds generally and is not specific to the initial product states of effective spins. 

\subsection{Multi-spin observables}\label{multispin}

Here we analyze time evolution of observables that involve two or more effective spins. We will show that, similar to the case of single-spin operators, such observables show power-law decay in time; however, there are different regimes characterized by different power-law exponents. We will use the results of this Subsection to analyze the physical initial states in the next Section. 

In order to understand the behaviour of expectation values of multi-spin operators, it is useful to go from 
the Schr\"odinger  to the Heisenberg representation, where operators, rather than wave functions, depend on time.  Using the simple form of the effective Hamiltonian, we first obtain the time-dependent form of the single-spin operators:
\begin{subequations}\label{eq:tau_t}
\begin{eqnarray}\label{eq:tau_t1}
\tau_k^z(t)&=&\tau_k^z,  \\
\label{eq:tau_t2}
\tau_k^x(t)&=&\cos(2\hat H_k t) \tau_k^x-\sin (2\hat H_kt) \tau_k^y, \\
\label{eq:tau_t3}
\tau_k^y(t)&=&\cos (2\hat H_k t) \tau_k^y+\sin(2\hat H_kt) \tau_k^x, 
\end{eqnarray}
\end{subequations}
where $\hat H_k$ is the magnetic field experienced by the spin $k$ [see Eq.~(\ref{eq:Heff})], which is an operator itself, although diagonal in the basis of effective spins. 

We note that using the expression for $\tau_k^\alpha (t)$ in the Heisenberg representation, Eq.~(\ref{eq:tau_t}), one can re-derive the above answers for $\corr{\tau^{x,y,z}_k (t)}$ [Eqs.~(\ref{eq:tauz})-(\ref{eq:taux})]. Obtaining the expectation values of multi-spin operators is also straightforward.  It is sufficient to understand the expectation value of an operator that is a product of several  $\tau_{i}^\alpha$ operators: 
\begin{equation}\label{eq:Tdef}
\hat T_{\{ i\}}^{\{\alpha \}}  = \hat T_{i_1 i_2\ldots i_n}^{\alpha_1 \alpha_2...\alpha_n}=\tau_{i_1}^{\alpha_1} \tau_{i_2}^{\alpha_2}...\tau_{i_n}^{\alpha_n},
\end{equation}
where $\alpha_{i}=x,y,z$. If $\alpha_i=z$ for all $i_1, i_2,..i_n$, from Eq.~(\ref{eq:tau_t1}) it follows that the expectation value of any string of $\tau^z$ at different sites remains constant with time,
\begin{equation} \label{eq:Tz}
\la \hat T_{i_1 i_2\ldots i_n}^{zz\ldots z}(t)\ra =\la \hat T_{i_1 i_2\ldots i_n}^{zz\ldots z}(0)\ra.
\end{equation}
On the other hand, the presence of $\tau^x$ or $\tau^y$ operators in Eq.~(\ref{eq:Tdef}) induces oscillations and decay in time as remote spins become entangled.

We illustrate this by analyzing the expectation value for the two-spin operator, $\hat T_{jk}^{xx} = \tau^x_{j}\tau^x_{k}$. Using Eq.(\ref{eq:tau_t}), we obtain: 
\begin{eqnarray}\label{eq:Txx}
\nonumber \corr{T_{jk}^{xx}(t)} 
=
\sum_{\tau''} P_{\{\tau''\}}\sum_{\tau^z_{j},\tau^z_{k}} A^*_{j \bar \tau^z_{j}} A^*_{k \bar \tau^z_{k}} A_{j \tau^z_{j}} A_{k \tau^z_{k}}
\\
\times
\exp \left[-i(E_{\tau_j^z \tau_k^z\{\tau''\}} - E_{\bar{\tau}_j^z \bar{\tau}_k^z\{\tau''\}} )t \right],
\end{eqnarray}
where $\{\tau''\}$ refers to the configuration of all spins excluding $j$th and $k$th ones, and $\bar \tau^z_{k}$ denotes an opposite spin from $\tau^z_{k}$ (if $\tau^z_{k}=\uparrow$, then $\bar \tau^z_{k}=\downarrow$). The probability $P_{\{ \tau'' \}}$ for the specific initial state considered here is $P_{\{ \tau'' \}}=\prod_{i\neq j, k} |A_{i \tau_i}|^2$. The phase factor can be expressed via effective magnetic field at sites $j$ and $k$,
\begin{multline}
E_{\tau_j^z \tau_k^z\{\tau''\}} -E_{\bar{\tau}_j^z \bar{\tau}_k^z\{\tau''\}}  = \\
2H_j ( \tau_k^z\{  \tau'' \}) \tau_j^z + 2H_k( \bar{\tau}_j^z\{ \tau'' \} )\tau_k^z.
\end{multline}
Note that terms proportional to $\tau^z_j \tau^z_k$ drop off from above equation. Indeed, operator $\tau_j^x\tau_k^x$ flips both spins simultaneously,  and the terms proportional to $\tau^z_j\tau^z_k$ are the same in $E_{\tau_j^z \tau_k^z\{\tau''\}}$ and $E_{\bar{\tau}_j^z \bar{\tau}_k^z\{\tau''\}}$.

The expectation value in Eq.~(\ref{eq:Txx}) is a sum of many oscillating terms, similar to the case of single-spin observables. However, the behaviour of this function at time $t$ depends on whether at that time the spins $j,k$ have become entangled or not. At times such that $t\ll J_0^{-1}e^{|j-k|/2\xi}$, the spins $j$ and $k$ evolve independently; in this regime, the sum in the r.-h.s.\ of Eq.~(\ref{eq:Txx}) can be represented a product of two sums, which are equal to $\la \tau_j^x(t)\ra$ and $\la \tau_k^x(t)\ra$, respectively. Using the results from previous Subsection, we obtain that $\la\tau_j^x (t) \tau_k^x (t) \ra \propto 1/t^{2b}$. At long times, $t\gg J_0^{-1}e^{|j-k|/\xi}$, when spins $j$ and $k$ are entangled, the sum in Eq.~(\ref{eq:Txx}) contains $2^{2l}$ independent random terms, where $l\sim \xi \log(J_0 t)$. 
Then, using an argument similar to the one from Section~\ref{singlespin}, we obtain an estimate $\corr{T_{jk}^{xx}(t)}\propto 1/t^b$. This behaviour can be summarized as follows: 
\begin{subequations}\label{eq:Txx_answer}
\begin{eqnarray}\label{eq:Txx_short}
\corr{T_{jk}^{xx}(t)} &\propto & \frac{1}{t^{2b}},\,\,  t\ll J_0^{-1}e^{|j-k|/\xi}  \\
\label{eq:Txx_long}
\corr{T_{jk}^{xx}(t)}& \propto &\frac{1}{t^{b}},\,\,  t\gg J_0^{-1}e^{|j-k|/\xi}. 
\end{eqnarray}
\end{subequations}
Combining results for two- and single-spin correlators, Eqs.~(\ref{eq:taux}) and (\ref{eq:Txx_answer}), we can understand the behaviour of the irreducible correlation function, defined as the difference
\begin{equation}\label{corrfun}
\corr{\tau^\sigma_j\tau^{\sigma'}_k}_c=\corr{\tau^\sigma_j\tau^{\sigma'}_k}-\corr{\tau^\sigma_j}\corr{\tau^{\sigma'_k}}.
\end{equation}
 At short times, $t\ll J_0^{-1}e^{|j-k|/2\xi}$, the irreducible correlation function $\la \tau_j^x (t) \tau_k^x (t) \ra_c$ is zero, and it saturates according to a power-law at long times, once the correlations between spins $j$ and $k$ have developed.

These considerations can be generalized for products of three and more spin operators. If the string $\hat T_{i_1 i_2...i_n}^{\alpha_1 \alpha_2...\alpha_n}$ contains at least one $\tau^x$ or $\tau^y$ operator, at sufficiently long times (such that $t\gg J_0^{-1} e^{-|s|/\xi}$, where $s=\max (|i_p-i_q|)$ is the support of the operator $\hat T_{i_1 i_2...i_n}$) the corresponding expectation value decays as a power-law with exponent $b$: 
\be\label{eq:manyspins}
\la \hat T_{i_1 i_2...i_n}^{\alpha_1 \alpha_2...\alpha_n} (t)\ra \propto \frac{1}{t^b}, \quad 
\text{if} \quad \exists \alpha_i \in \{x,y\}.  
\ee
At shorter times, this function can decay with a different power-law, similar to the case of $\la T^{xx}_{jk} (t) \ra$. We will use the result Eq.~(\ref{eq:manyspins}) in the following Section to understand the behaviour of physical operators. 

\subsection{Higher dimensions and strongly entangled initial states \label{sec-comment}}

Before we address physical observables, let us briefly comment on the extension of previous results to higher dimensions or more complicated initial states. Assuming the existence of an infinite-temparature MBL phase in higher dimensions, it is straightforward to repeat the derivation of the off-diagonal element of the density matrix, Eqs.~(\ref{eq:estimate})-(\ref{eq:estimate2}). In $d$ dimensions, entanglement spreads over a sphere, hence the number of oscillating terms in $\rho_{\uparrow\downarrow}(t)$ grows as $N(t) = 2^{V_d l^d}$, where $V_d$ is the volume of a unit sphere in $d$ spatial dimensions. Assuming relation $l\sim \xi \ln (J_0 t)$ holds, we arrive at a faster-than-power-law decay of correlations,  
\begin{equation} \label{Eq:rho-higher}
|\rho_{\uparrow\downarrow}(t)| \sim \frac{|A_{k\uparrow}A_{k\downarrow}^*|}{(tJ_0)^{a' [\ln (tJ_0)]^{d-1} }},
\end{equation}
where $a' = (\ln 2/2) V_d \xi^d$ when initial state has maximum Renyi entropy.

Another interesting question pertains to a quench from the initial state that is strongly entangled, i.e., violates the boundary-law. Experimentally such initial states can be realized, for example, by abruptly changing disorder strength in the model~(\ref{Eq:XXZ}), thus tuning the system from the delocalized into the MBL phase. Returning to the general result for $\rho_{\uparrow\downarrow}(t)$,  Eq.~(\ref{eq:reduced2-1}), we see that strongly entangled initial states will result in a more complicated structure of probability $P_{\{\tau'\}}$. However, provided there is an extensive entropy density (that is, the number of terms contributing to sum in Eq.~(\ref{eq:reduced2-1}) grows exponentially with the system size), logarithmic spreading of entanglement results in a slow power-law decay of observables to their long-time values.

\section{Global quench: physical observables \label{sec:general}}

Now we consider physical spin observables, and argue that their behaviour is similar to the effective spin operators considered above: the average values reach ``equilibrium'' (but non-thermal) values at long times, which depend on the initial state. The fluctuations around these values decay in a power-law fashion as a function time. Also, we use numerical simulations of the random-field $XXZ$ spin chain to support our analytic arguments. 

\subsection{Analytic considerations}

We assume that the system is prepared in some initial state $|\Psi_0\ra$ at $t=0$, which can be expanded in the basis of eigenstates as follows: 
$$
|\Psi_0\ra=\sum_{\{ \tau\}} A_{\{ \tau \}} |\{\tau \}\ra, 
$$
where $\{\tau \}=\tau_1^z \tau_2^z...\tau_L^z$, $\tau_i^z=\uparrow,\downarrow$ denotes $2^L$ eigenstates of our system. We focus on the experimentally relevant class of initial states which are product states of physical spins. However, we expect our analysis to apply to a more general class of initial states, in particular those obeying the boundary-law entanglement. 

We will be interested in the time evolution of a local observable described by the local operator $\hat{\mathcal{O}}$ acting on several physical spins situated near site $k$. Such an operator can be expanded in the basis of the effective spin operators (\ref{eq:Tdef}):  
\be\label{eq:local_operator}
\hat{\mathcal{O}}=\sum_{\{i\}, \{ \alpha\} } B_{\{ i \}}^{ \{\alpha\}} \hat T_{\{ i\}}^{\{\alpha\}}, 
\ee
where $\{ i\}$ runs over groups of effective spins, and $\{ \alpha \}$ denotes various combinations of spin projections for a given choice of $i$. For example, for the case $\hat{\mathcal O}=\sigma_k^z$,
\begin{equation} \label{Eq:sigmaz-exp}
\sigma^z_k =\sum_{i_1}B^{z}_{i_1} \tau^z_{i_1} + \sum_{i_1,i_2,\sigma=x,y,z} B^{\sigma\sigma}_{i_1i_2} \tau^\sigma_{i_1}\tau^\sigma_{i_2} +\ldots,
\end{equation}
where ellipses denote three- and  higher-order spin terms. In writing the above equation, we assumed that $\tau$-operators are chosen such that $\sigma_k^z$ commutes with the total $\tau^z$ operator (for a model with conserved total $s^z$ such a choice is always possible), and therefore terms such as $\tau_{i_1}^{x,y}$ are not allowed on the r.-h.s. of Eq.~(\ref{Eq:sigmaz-exp}).  

Due to the locality of the operator $\hat{\mathcal{O}}$ and the quasi-locality of the unitary transformation that relates physical and effective spins, coefficients $B_{\{i \}}^{\{ \alpha \}}$ decay exponentially  for larger groups of effective spins. Presumably,
$$
B_{\{i \}}^{\{ \alpha \}}\propto \exp(-l/\chi), 
$$
where $l={\rm max}(|i_p-k|)$ is the range of the operator $\hat T_{\{ i\}}^{\{\alpha\}}$, and  $\chi$ is set by the properties of the quasi-local unitary transformation that diagonalizes the MBL Hamiltonian.  

In order to understand the time evolution of the expectation value of operators $T_{\{ i\}}^{\{\alpha\}}$ for initial product state $|\Psi_0\ra$, we note that the latter is a superposition of (exponentially) many eigenstates $\{ \tau\}$. Phases $H_k t$ become random at long times for wave function components with different $\{ \tau' \}$, which leads to the suppression of
the averages of "off-diagonal" operators $T_{\{ k\}}^{\{\alpha\}}$ (the ones where at least one $\alpha_i=x,y$). An argument similar to the one described in the previous Section then shows that each $\la \Psi_0| T_{\{ k\}}^{\{\alpha\}} |\Psi_0\ra$ in sum~(\ref{eq:local_operator}) decays as a power-law function of time. 

Turning to physical observables, we note that the operator $\hat{\mathcal{O}}$ can be represented as a sum of its part that commutes with the Hamiltonian, $\bar{\mathcal O}$, and the remaining part. The commuting part  is obtained by taking terms in the expansion (\ref{eq:local_operator}) in which all $\alpha_i=z$, and the remainder is given by the sum of terms where at least one index $\alpha_i=x, y$.  The conserved part $\hat{\mathcal O}$ is a local integral of motion (see Ref.~[\onlinecite{Chandran14}] for a detailed study of such operators). It does not change under time evolution, and retains the memory of the initial state. 

The ``off-diagonal'' part, being a sum of a finite number of ``off-diagonal'' operators $T_{\{ i\}}^{\{\alpha\}}$, exhibits oscillations with a typical magnitude that decays as a power-law in time at sufficiently long times (see Eq.(\ref{eq:manyspins})). This can be summarized as
\begin{subequations}\label{eq:phys-corr}
\begin{eqnarray}\label{eq:long-time}
\la \hat{\mathcal O} (t)  \ra &\to& \la  \bar {\mathcal O} \ra  ,\quad t\to \infty 
\\
\label{eq:long-time_fluct}
\big(\la \hat{\mathcal O} (t) \ra - \la \bar{\mathcal O} \ra\big)^2 &\sim& \frac {1}{t^{2b}},\quad t\gg 1/J_0,  
\end{eqnarray}
\end{subequations}
{where expectation value is taken with respect to the state $|\Psi_0\ra$}.

Let us comment on the applicability of the above results, Eq.~(\ref{eq:phys-corr}). In Sec.~\ref{sec-comment} we argued that power-law decay is not limited to initial product states, but rather holds for any initial states with finite entropy density, including highly entangled initial states. Thus we conclude that this behaviour applies to any \emph{local} observable in the MBL phase unless the initial state is fine tuned and is very close to an eigenstate.  The power-law exponent $b$ is proportional to $\xi$, with a prefactor that generally depends on the way the initial state is prepared. This prefactor is set by the entropy density of the initial state in the effective spin basis. Thus, generally, $b\leq a = \xi\ln 2$, as $\ln 2$ is the maximum possible entropy density. This inequality becomes an equality if $|\Psi_0\ra$ is an equal-weight superposition of all the eigenstates, as  was the case for the initial state considered in Sec.~\ref{Sec:quench}.

As we argued above, slow, power-law-like decay also holds for longer-range observables, e.g. correlation function between two distant sites $j$ and $k$, $\corr{\sigma^x_j \sigma^x_k}$. However, the exponent of the decay in Eq.~(\ref{eq:Txx_answer}) is non-universal, as it changes on the time scale over which correlations between spins $j$ and $k$ develop. In principle, all physical operators, according to Eq.~(\ref{eq:local_operator}), contain terms that are long ranged but exponentially suppressed. However, as we demonstrate below numerically, these terms do not matter for the simplest local operators, which exhibit clear power-law behaviour.

\begin{figure}[tbh]
\begin{center}
\includegraphics[width=0.99\columnwidth]{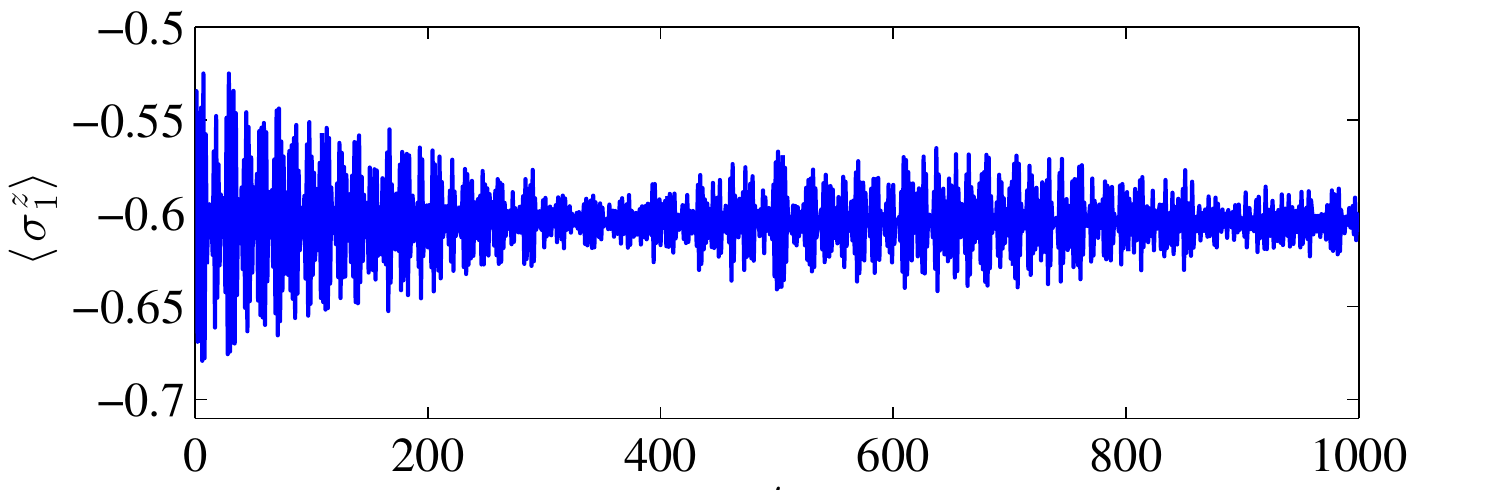}
\setlength{\unitlength}{\columnwidth}
\begin{picture}(0,0)
\put(-0.5,0.41){(a)}
\put(-0.5,-0.01){(b)}
\put(-0.5,-0.41){(c)}
\put(-0.0,0.02){$t$}
\end{picture}
\includegraphics[width=0.99\columnwidth]{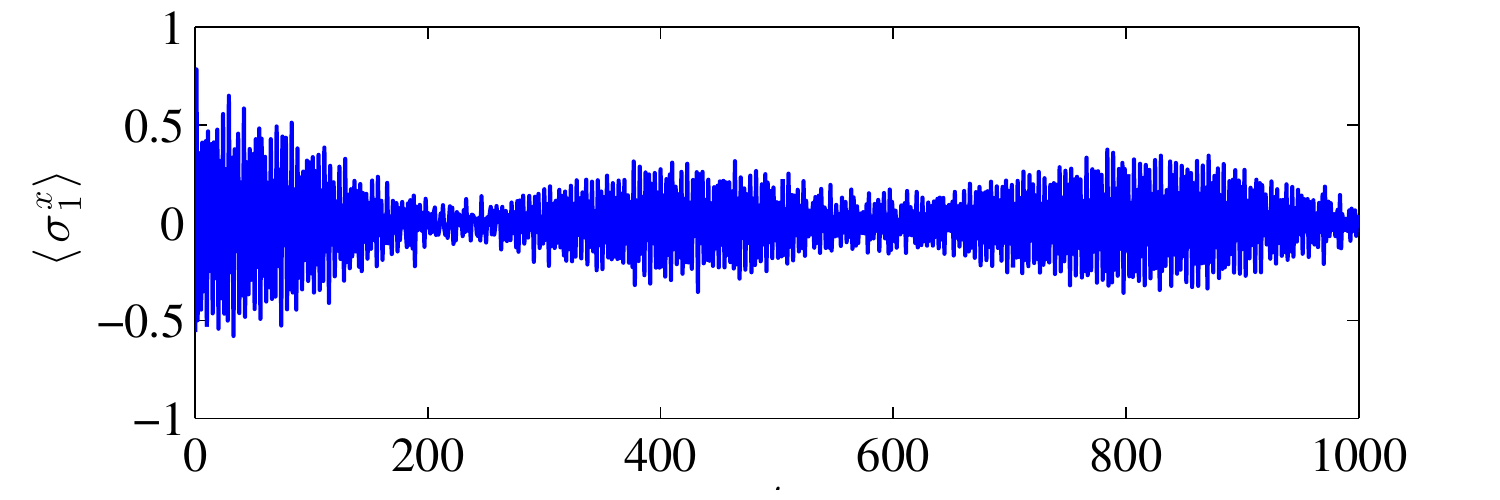}
\setlength{\unitlength}{\columnwidth}
\begin{picture}(0,0)
\put(0.0,0.02){$t$}
\end{picture}
\includegraphics[width=0.99\columnwidth]{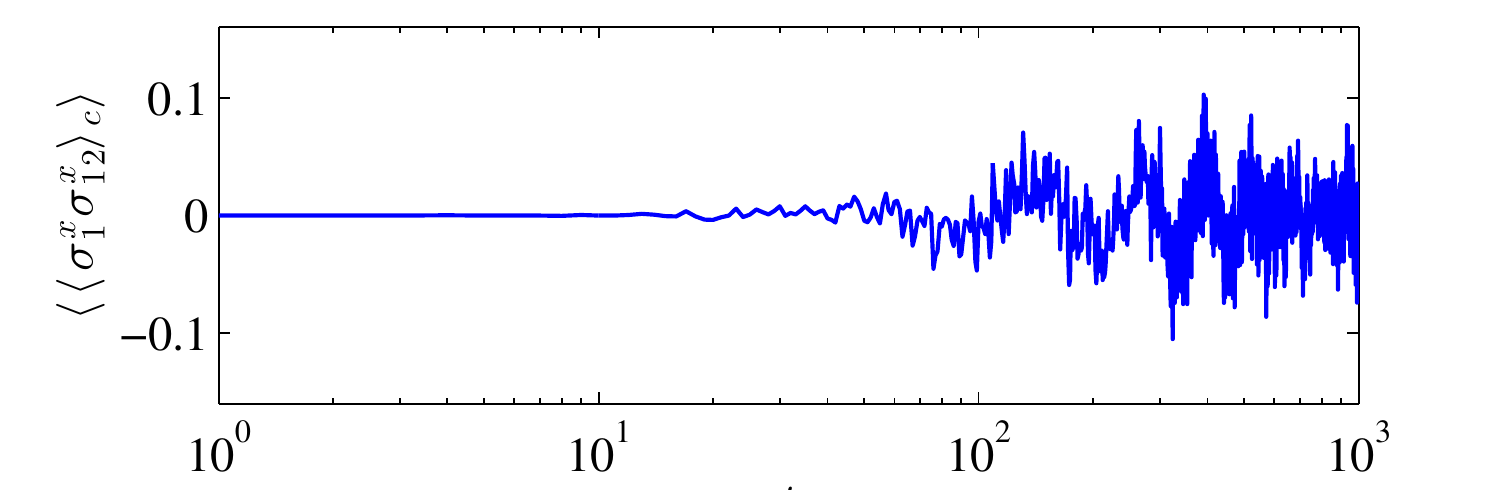}
\setlength{\unitlength}{\columnwidth}
\begin{picture}(0,0)
\put(0.0,0.02){$t$}
\end{picture}
\caption{ \label{Fig:relax_noav} Expectation values of local operators and correlation function for an initial state where each spin points in a random direction. Panels (a) and (b) show the expectation value of local operators $\sigma^z_1$ and $\sigma^x_1$. The signal is oscillating at many different frequencies. While the envelope indeed decays, one can see multiple revivals of the signal. The irreducible spin-spin correlation function, panel (c), is zero at small times, with correlations developing at larger times. No disorder or ensemble averaging were performed.  Interactions $J_z=1$, disorder $W=5$, system size is $L=12$.}
\end{center}
\end{figure}
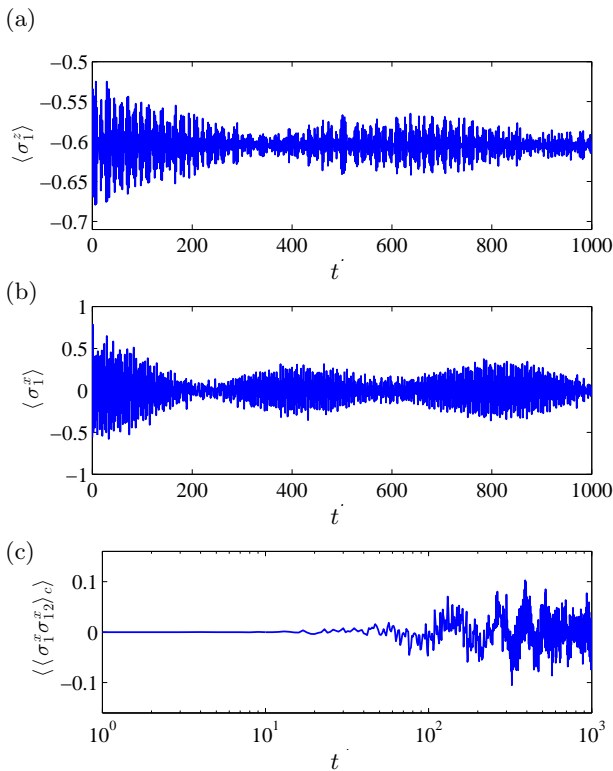

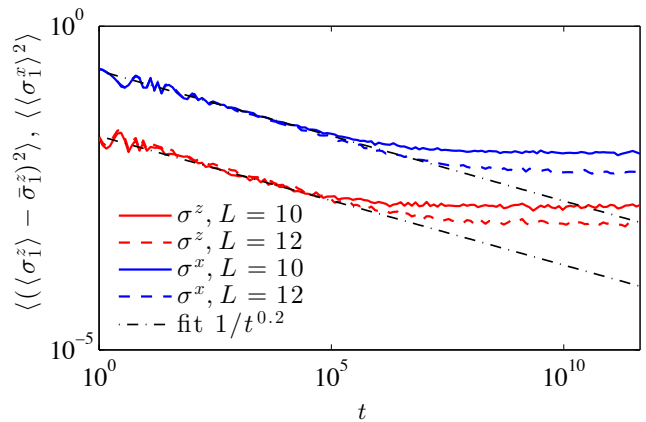
\begin{figure}[tbh]
\begin{center}
\includegraphics[width=0.99\columnwidth]{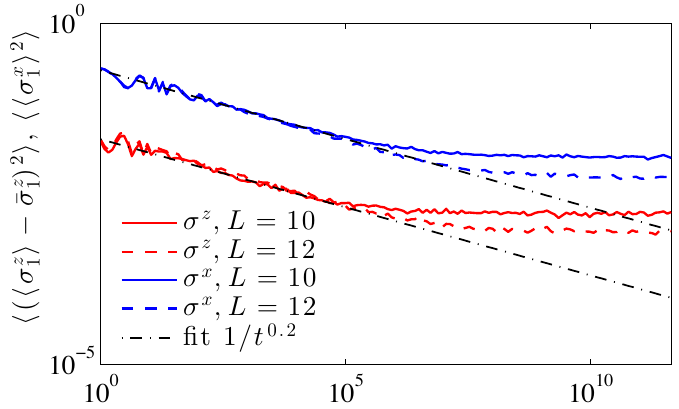}
\setlength{\unitlength}{\columnwidth}
\begin{picture}(0,0)
\put(0.05,0){$t$}
\end{picture}
\caption{ \label{Fig:relax_dav} Disorder-averaged expectation values of local operators $\sigma^z_1$ and $\sigma^x_1$. The fluctuations around the mean value decay in a power-law fashion with the same exponent that does not depend on system size. However, the saturation value decreases with system size. The initial states are chosen from an ensemble of states for which each spin initially points in a random direction. Interactions $J_z=1$, disorder $W=5$, system size is $L=10$ and $12$. Averaging was performed over $4000$  configurations.}
\end{center}
\end{figure}

\subsection{Numerical tests\label{Sec:sim}}

\begin{figure}
\begin{center}
\includegraphics[width=0.99\columnwidth]{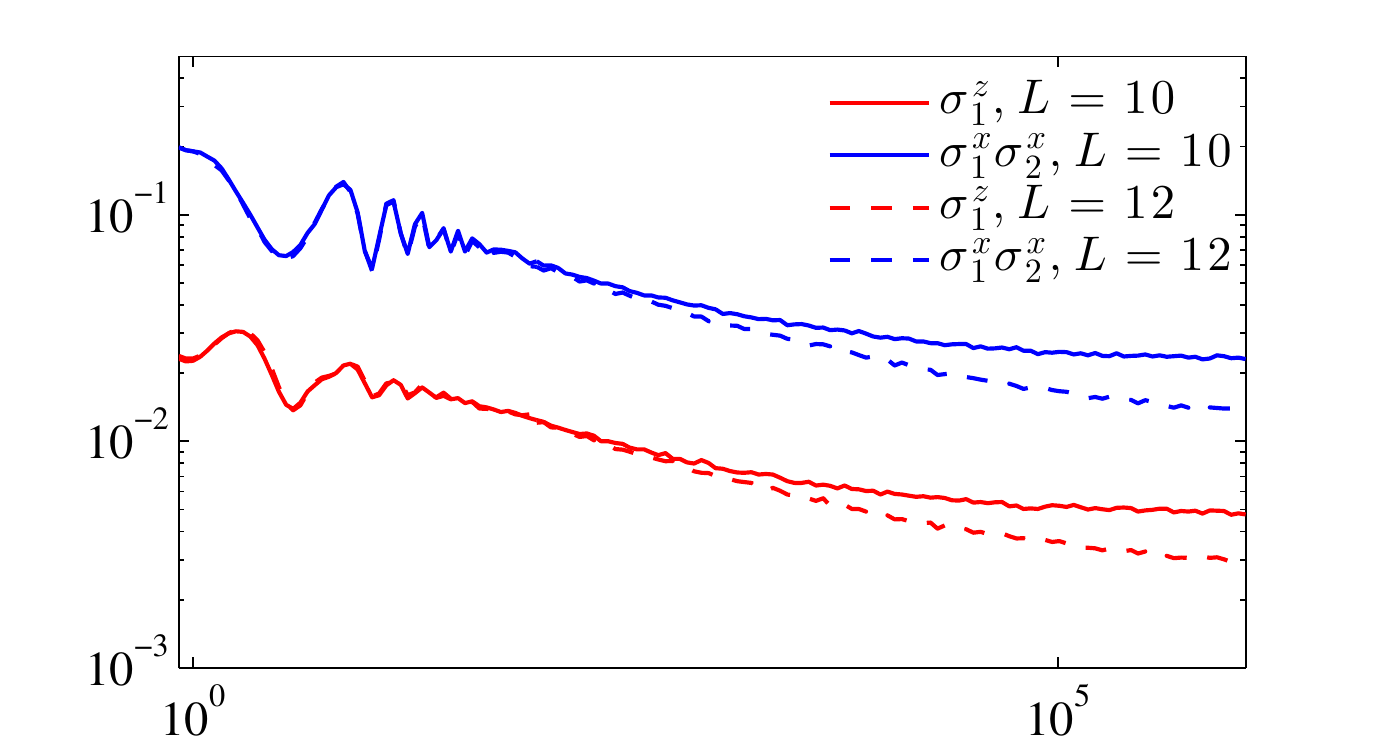}
\setlength{\unitlength}{\columnwidth}
\begin{picture}(0,0)
\put(-0.47,0.18){\rotatebox{90}{$\color{blue}\langle(\langle \sigma_1^x\sigma_{2}^x \rangle- {\overline{ \sigma_1^x\sigma_{2}^x}})^2\rangle$}}
\put(-0.51,0.2){\rotatebox{90}{$\color{red} \langle(\langle \sigma_1^z\rangle-\bar\sigma_1^z)^2\rangle,$}}
\put(-0.0,0.02){$t$}
\end{picture}
\caption{ \label{Fig:str_ent} Dynamics of local operators when the initial state is the ground state of XXZ spin chain with vanishing disorder $W=0.05$. Interactions $J_z=1$, disorder $W=5$, system size is $L=10$ and $12$. Averaging was performed over $4000$  configurations.}
\end{center}
\end{figure}
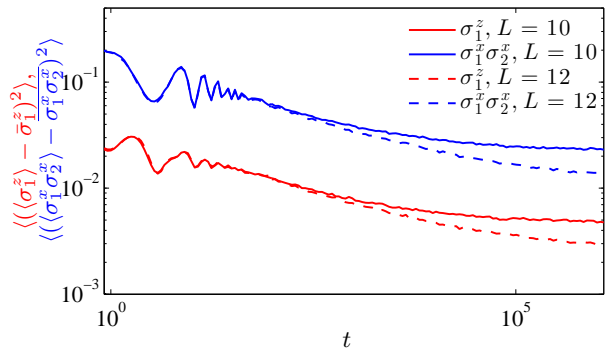

\begin{figure}[tbh]
\begin{center}
\includegraphics[width=0.99\columnwidth]{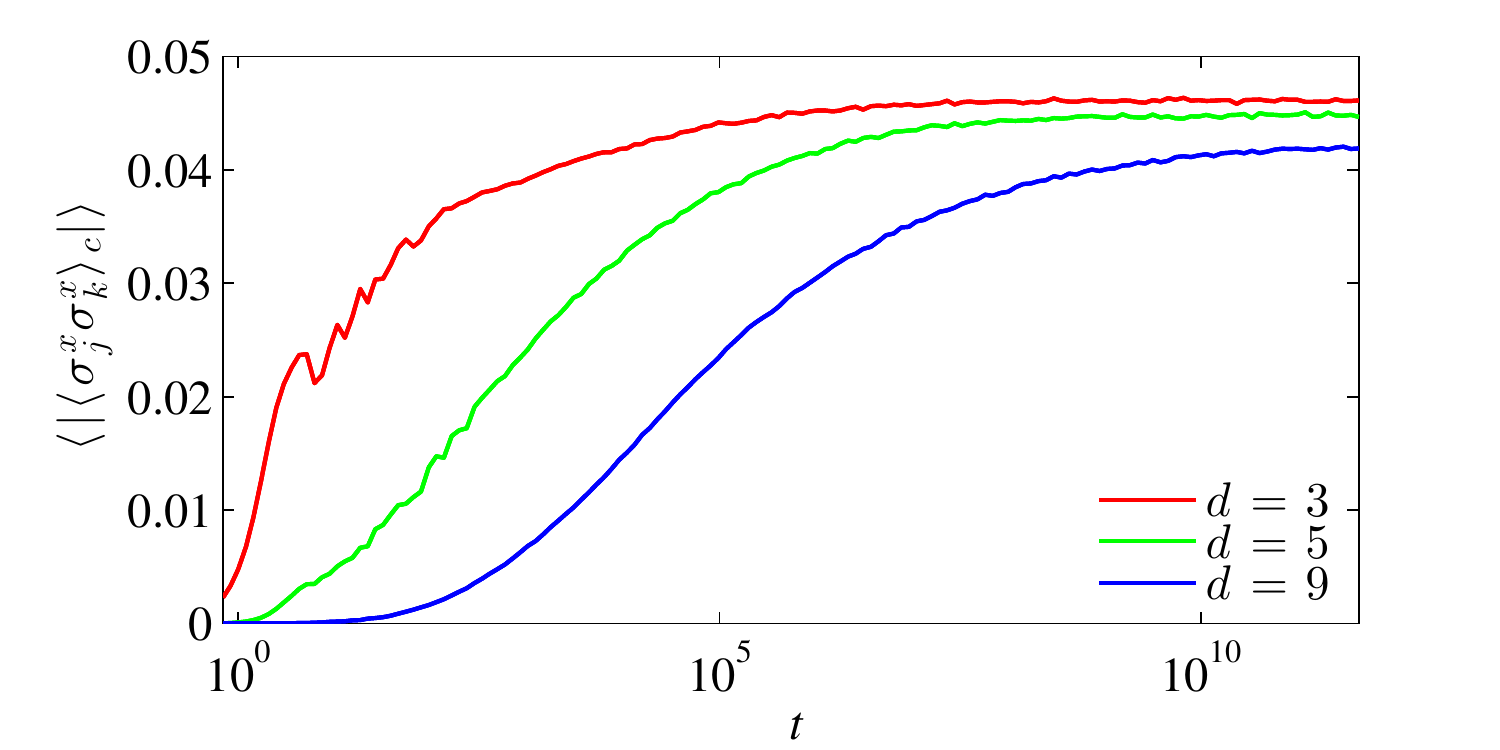}
\caption{ \label{Fig:relax_xxav} Disorder-averaged expectation values of the irreducible correlation function as a function of distance between sites $j$ and $k$ for a chain of $L=10$ spins. The growth of correlation is exponentially sensitive to the distance between spins, however it has the same slope.  The initial states are chosen from an ensemble of states for which each spin initially points in a random direction. Interactions $J_z=1$, disorder $W=9$. }
\end{center}
\end{figure}

To test our analytic results for the behaviour of local observables and correlation functions, here we  study the quench dynamics in the MBL phase numerically. We focus on the XXZ spin chain, described in detail in Sec.~\ref{Sec:modelXXZ}, as a model for the MBL phase. 

We employ exact diagonalization to study time evolution and local observables of the Hamiltonian~(\ref{Eq:XXZ}). We consider spin chains of size $L=10$ and $L=12$ with open boundary conditions, without restricting to a particular spin sector of the Hilbert space.   First, we prepare the initial state as a product state, where each physical spin, $\sigma_{i}$, $i=1,\ldots, L$ points in a random direction on the Bloch sphere.  Dynamics of local {spin operators on the very first site}, $\corr{\sigma^z_1}$, $\corr{\sigma^x_1}$, along with correlation function~$\langle \sigma^x_1 \sigma^x_L \rangle_c$ for a single such initial state is shown in Fig.~\ref{Fig:relax_noav}. For local observables we see fast oscillations on the scale set by $J_\perp$ and local magnetic field $w_i$. While the amplitude of these oscillations indeed decays, clear revivals of the signal are present. A notable difference between $\sigma^x$ and $\sigma^z$ operators is that the latter oscillates around a non-zero value~[Fig.~\ref{Fig:relax_noav}(a)], indicating that this operator has a non-vanishing overlap with the conserved quantity $\tau_1^z$. 

The irreducible correlation function (\ref{corrfun}) between physical spins at the opposite ends of the chain, $\corr{\sigma^x_1\sigma^x_L}_c$~[Fig.~\ref{Fig:relax_noav}(c)], behaves differently from local observables. It is strictly zero at short times, since we start from a product state of physical spins. At later times, the correlation function becomes non-zero and starts to oscillate. Note, that   $\corr{\sigma^x_1\sigma^x_L}_c$ deviates from zero on a relatively short time scale. This is related to the presence of exponential tails in the relation between physical and effective spins.

In order to reveal the power-law decay of local observables and the logarithmic propagation of correlations, we need to suppress the fast oscillations in the signal illustrated in Fig.~\ref{Fig:relax_noav}. The simplest route to this is to consider the absolute value or square of the correlation function, and average it over disorder realizations. Fig.~\ref{Fig:relax_dav} shows results for the square of the spin expectation value, where averaging is performed over random disorder realizations and random spin directions in the initial state. Note that when we consider the square of spin $\sigma^x$ correlation function,  $\corr{\corr{\sigma^x_1}^2}$, we subtract off the long-time average of the~$\sigma^z$ correlator, $\corr{(\corr{\sigma^z_1}-\la\bar\sigma^z_1\ra)^2}$, to reveal how $\langle \sigma^x \rangle^2$ vanishes at long times. We see from  Fig.~\ref{Fig:relax_dav} that both averages decay in a power-law fashion with time, with approximately equal exponents. We note that the system size and explicit form of the operator ($\sigma^x$ or $\sigma^z$) controls the saturation value and time (compare solid and dashed line in Fig.~\ref{Fig:relax_dav}), but does not influence the exponent of decay. 

We also test the power-law decay for a different class of initial states that are not product states. Fig.~\ref{Fig:str_ent} displays the dynamics of local operators when the  system is initially prepared in the ground state of the Hamiltonian~(\ref{Eq:XXZ}) with $W=0.05$ (delocalized phase), and at $t=0$ disorder is abruptly changed to $W=5$. Note that saturation sets in much faster compared to the case of initial product states, Fig.~\ref{Fig:relax_dav}, despite the interaction and disorder strength being the same. 

Finally, we compute the averaged absolute value of the irreducible correlation function~$\corr{|\corr{\sigma^x_1\sigma^x_L}_c|}$ in Fig.~\ref{Fig:relax_xxav}.  The slope of the growth is not influenced by the distance between measured spins. Moreover, the increase in the distance between spins causes an exponential delay in the development of correlations. For the shortest distance between spins, one observes ballistic development of correlations at short times, followed by a slow growth [red curve in Fig.~\ref{Fig:relax_xxav}]. Note that although the irreducible correlation function also displays slow dynamics, it cannot be fitted with a simple power law. This reflects the fact that the decay of two-spin observables is described by power-laws with different exponents at short and long times, as discussed above.

\section{Local quench \label{Sec:quench2}}

In this Section, we propose an alternative setup for probing the characteristic dynamics of the MBL phase, which we refer to as a local quench. We assume that an isolated system as a whole is initially prepared in a thermal state (e.g., because it was brought in contact with an external bath and then disconnected from it). We study a setup in which some test spin is repeatedly prepared in the same quantum state, and then its time evolution is probed. If the test spin does not interact with the remaining degrees of freedom, it will generically exhibit non-decaying oscillations. However, residual interactions with an MBL system (which is in a mixed state) lead to dephasing and power-law decay of oscillations, similar to the case of the global quench studied above. 

Consider first a simplified situation when a test {\it effective} spin $k$ (rather than the physical spin) is prepared in the state $A_{k\uparrow}|\!\!\uparrow\ra+A_{k\downarrow}|\!\!\downarrow\ra$ at time $t=0$, while the remaining spins are in a mixed state, with the probability of state $\{ \tau'\}$ being $P_{\{ \tau'\}}$. The time evolution of the reduced density matrix of spin $k$ is given by Eqs.(\ref{eq:reduced1})-(\ref{eq:reduced2}), which we obtained for the case of a global quench. The physical observables therefore behave as above, with $\tau_k^z$ being conserved during evolution, and $\tau_k^{x,y}$ displaying power-law decay to zero at long times. The dephasing responsible for the power-law decay of $\tau_k^{x,y}$ in this case arises because the system is in a mixed state, and spin $k$ precesses with different frequency depending on the state of the remaining spins.   

In a realistic experiment, one would manipulate physical, rather than effective spins, and prepare spin $k$ in some initial state $A_{k\uparrow}|\!\!\uparrow\ra+A_{k\downarrow}|\!\!\downarrow\ra$. The physical spin $k$ is not a precise integral of motion, and therefore its $\sigma_k^z$ projection is not conserved. However, the operator $\sigma_k^z$ is expected to have a non-zero overlap with the integral of motion $\tau_k^z$ (see, e.g., Ref.~[\onlinecite{Serbyn14}]), thus, $\sigma_k^z$ will typically remain finite at long times, if for the initial state $\la \sigma_k^z\ra\neq 0$. Further, applying the arguments of the previous Section, we conclude that $\sigma_k^z$, as well as $\sigma_k^{x,y}$ approach their long-time values in a power-law fashion. Thus, a local quench provides an alternative way of experimentally probing dephasing dynamics that characterizes the MBL phase.

The situation changes if the MBL system is prepared in an eigenstate, rather than in a mixed state. In this case, the dephasing is suppressed, and the test spin will show non-decaying oscillations with a frequency that depends on the state of remaining spins. 

Although it is difficult to prepare the whole MBL system in an eigenstate, it should be possible to control the state of several spins around the test one. As we now argue, this ability can potentially be used to suppress dephasing and create long-lived quantum states of the test spin. Let us assume that the effective spins distance $r\gg \xi$
 away from the test spin can be prepared in states $|\!\!\uparrow\ra$ or $|\!\!\downarrow\ra$. We will call this region a "buffer region". In a random-field XXZ model, this can be achieved by polarizing all spins in the buffer region along the same (up or down) direction. Being in an eigenstate, these spins cannot get entangled and dephase spin $k$ -- their only effect is to change the precession frequency of spin $k$. Therefore, the test spin $k$ will only dephase and lose coherence after time of order $t_{\rm deph}\sim J_0^{-1} \exp (-r/\xi)$, when the interactions with the spins outside "buffer" become important. This provides a way of protecting the quantum states of the test spin, which complements spin-echo techniques.~\cite{Serbyn14}

 \section{Conclusions \label{Sec:concl}}
 
In conclusion, we studied the dynamics in the MBL phase following a quantum quench. 
We demonstrated that local observables reach stationary values at long times as a result of slow dephasing between remote conserved degrees of freedom, characteristic of the MBL phase~\cite{Serbyn13-1,Serbyn13-2,HuseOganesyan13,Altman13}. 
The steady state is highly non-thermal, and the memory of the initial state is contained in the values of local integrals of motion. For the local operators that represent the density of some extensive conserved quantity (such as the $z$-axis spin projection, or energy density), the long-time value is generally non-zero and is correlated with the initial value. More generally, this holds for local operators that have non-zero overlap with a local integral of motion in the MBL phase. We note that the existence of a stationary state distinguishes the MBL phase from a non-interacting Anderson insulator, where local observables show non-decaying oscillations following a quench. 

We argued that the time evolution of local observables, and their approach to the stationary values is also universal -- the fluctuations around the stationary values decay according to a power-law in time. The exponent is set by the localization length and  the entropy of the initial state. The power-law decay of local observables stems from the same mechanism that underlies the logarithmic in time growth of entanglement in the MBL phase~\cite{Serbyn13-1,HuseOganesyan13}; however, it has the advantage of being experimentally measurable. 
Further, we argued that an alternative probe of the slow spreading of information in the MBL phase is given by the correlation functions between remote degrees of freedom: the correlations start developing only at times which are exponential in the separation between those degrees of freedom. This is in contrast with ballistic spreading of correlations in ergodic or integrable systems, and  is described by the so-called zero-velocity Lieb Robinson bound.~\cite{Hamza,Kim-prep}

Altogether, the properties described above comprise a new regime of dynamics that arises in the MBL phase. In particular, it is instructive to contrast the MBL case with the case of ergodic systems. In the latter case, the system reaches an equilibrium thermal state at long times, with properties determined only by the globally conserved quantities (e.g., energy). Local observables typically approach their thermal values in an exponential fashion, except for local densities of the globally conserved quantities, which propagate diffusively and show power-law relaxation. 

Our article complements the related works~\cite{Bahri,Serbyn14, Nandkishore14, Johri14, Vasseur14} which discuss dynamical experimental signatures of the MBL phase. The modified spin-echo protocol, introduced in Ref.~[\onlinecite{Serbyn14}], probes the dephasing of a given spin due to its interaction with a specific remote region. An alternative route, suggested in Ref.~[\onlinecite{Vasseur14}], is to probe the dephasing in the MBL phase by measuring the revivals in the magnetization of a test qubit coupled to a long chain. The universal power-law relaxation, identified in the present work, can be observed in a natural setup involving a global or local quench, and is also robust to thermal and disorder averaging, similar to the modified spin-echo protocol~\cite{Serbyn14}. It is worth mentioning that spectral properties of the MBL phase, providing additional experimental signatures, were considered in Refs.~[\onlinecite{Nandkishore14},\onlinecite{Johri14}]. 

Finally, we note that the setup considered in this paper can be realized in systems of cold atoms in optical lattices, where both disorder and interactions can be controlled in a broad range. For example, the behaviour of local observables can be studied in a disordered Bose-Hubbard or Fermi-Hubbard model in one dimensional optical lattice, by preparing a non-uniform initial state~\cite{DeMarco10,Inguscio13,DeMarco13}. In the MBL phase, local density modulation will remain finite at long times. Further, MBL phase can be detected by the characteristic power-law behaviour of local observables, as well as by measuring the time evolution of correlation functions. The light cone spreading of correlations in the ergodic phase has been recently observed experimentally for a one dimensional quantum gas~\cite{Bloch12}. In closing, we note that it would be interesting to study quantum quenches at and near the transition between MBL and ergodic phase.

\section*{Acknowledgements}

Research at Perimeter Institute is supported by the Government of Canada through Industry
Canada and by the Province of Ontario through the Ministry of Economic Development \& Innovation.
We acknowledge support by NSERC Discovery Grant (D.A.).


\begin{thebibliography}{44}%
\makeatletter
\providecommand \@ifxundefined [1]{%
 \@ifx{#1\undefined}
}%
\providecommand \@ifnum [1]{%
 \ifnum #1\expandafter \@firstoftwo
 \else \expandafter \@secondoftwo
 \fi
}%
\providecommand \@ifx [1]{%
 \ifx #1\expandafter \@firstoftwo
 \else \expandafter \@secondoftwo
 \fi
}%
\providecommand \natexlab [1]{#1}%
\providecommand \enquote  [1]{``#1''}%
\providecommand \bibnamefont  [1]{#1}%
\providecommand \bibfnamefont [1]{#1}%
\providecommand \citenamefont [1]{#1}%
\providecommand \href@noop [0]{\@secondoftwo}%
\providecommand \href [0]{\begingroup \@sanitize@url \@href}%
\providecommand \@href[1]{\@@startlink{#1}\@@href}%
\providecommand \@@href[1]{\endgroup#1\@@endlink}%
\providecommand \@sanitize@url [0]{\catcode `\\12\catcode `\$12\catcode
  `\&12\catcode `\#12\catcode `\^12\catcode `\_12\catcode `\%12\relax}%
\providecommand \@@startlink[1]{}%
\providecommand \@@endlink[0]{}%
\providecommand \url  [0]{\begingroup\@sanitize@url \@url }%
\providecommand \@url [1]{\endgroup\@href {#1}{\urlprefix }}%
\providecommand \urlprefix  [0]{URL }%
\providecommand \Eprint [0]{\href }%
\providecommand \doibase [0]{http://dx.doi.org/}%
\providecommand \selectlanguage [0]{\@gobble}%
\providecommand \bibinfo  [0]{\@secondoftwo}%
\providecommand \bibfield  [0]{\@secondoftwo}%
\providecommand \translation [1]{[#1]}%
\providecommand \BibitemOpen [0]{}%
\providecommand \bibitemStop [0]{}%
\providecommand \bibitemNoStop [0]{.\EOS\space}%
\providecommand \EOS [0]{\spacefactor3000\relax}%
\providecommand \BibitemShut  [1]{\csname bibitem#1\endcsname}%
\let\auto@bib@innerbib\@empty
\bibitem [{\citenamefont {Polkovnikov}\ \emph {et~al.}(2011)\citenamefont
  {Polkovnikov}, \citenamefont {Sengupta}, \citenamefont {Silva},\ and\
  \citenamefont {Vengalattore}}]{PolkovnikovColdAtoms}%
  \BibitemOpen
  \bibfield  {author} {\bibinfo {author} {\bibfnamefont {A.}~\bibnamefont
  {Polkovnikov}}, \bibinfo {author} {\bibfnamefont {K.}~\bibnamefont
  {Sengupta}}, \bibinfo {author} {\bibfnamefont {A.}~\bibnamefont {Silva}}, \
  and\ \bibinfo {author} {\bibfnamefont {M.}~\bibnamefont {Vengalattore}},\
  }\href {\doibase 10.1103/RevModPhys.83.863} {\bibfield  {journal} {\bibinfo
  {journal} {Rev. Mod. Phys.}\ }\textbf {\bibinfo {volume} {83}},\ \bibinfo
  {pages} {863} (\bibinfo {year} {2011})}\BibitemShut {NoStop}%
\bibitem [{\citenamefont {Bloch}\ \emph {et~al.}(2008)\citenamefont {Bloch},
  \citenamefont {Dalibard},\ and\ \citenamefont {Zwerger}}]{BlochColdAtoms}%
  \BibitemOpen
  \bibfield  {author} {\bibinfo {author} {\bibfnamefont {I.}~\bibnamefont
  {Bloch}}, \bibinfo {author} {\bibfnamefont {J.}~\bibnamefont {Dalibard}}, \
  and\ \bibinfo {author} {\bibfnamefont {W.}~\bibnamefont {Zwerger}},\ }\href
  {\doibase 10.1103/RevModPhys.80.885} {\bibfield  {journal} {\bibinfo
  {journal} {Rev. Mod. Phys.}\ }\textbf {\bibinfo {volume} {80}},\ \bibinfo
  {pages} {885} (\bibinfo {year} {2008})}\BibitemShut {NoStop}%
\bibitem [{\citenamefont {Blatt}\ and\ \citenamefont {Roos}(2012)}]{Blatt12}%
  \BibitemOpen
  \bibfield  {author} {\bibinfo {author} {\bibfnamefont {R.}~\bibnamefont
  {Blatt}}\ and\ \bibinfo {author} {\bibfnamefont {C.~F.}\ \bibnamefont
  {Roos}},\ }\href@noop {} {\bibfield  {journal} {\bibinfo  {journal} {Nature
  Physics}\ }\textbf {\bibinfo {volume} {8}},\ \bibinfo {pages} {277} (\bibinfo
  {year} {2012})}\BibitemShut {NoStop}%
\bibitem [{\citenamefont {Devoret}\ and\ \citenamefont
  {Schoelkopf}(2013)}]{Devoret13}%
  \BibitemOpen
  \bibfield  {author} {\bibinfo {author} {\bibfnamefont {M.~H.}\ \bibnamefont
  {Devoret}}\ and\ \bibinfo {author} {\bibfnamefont {R.~J.}\ \bibnamefont
  {Schoelkopf}},\ }\href@noop {} {\bibfield  {journal} {\bibinfo  {journal}
  {Science}\ }\textbf {\bibinfo {volume} {339}},\ \bibinfo {pages} {1169}
  (\bibinfo {year} {2013})}\BibitemShut {NoStop}%
\bibitem [{\citenamefont {Rigol}\ \emph {et~al.}(2008)\citenamefont {Rigol},
  \citenamefont {Dunjko},\ and\ \citenamefont {Olshanii}}]{RigolNature}%
  \BibitemOpen
  \bibfield  {author} {\bibinfo {author} {\bibfnamefont {M.}~\bibnamefont
  {Rigol}}, \bibinfo {author} {\bibfnamefont {V.}~\bibnamefont {Dunjko}}, \
  and\ \bibinfo {author} {\bibfnamefont {M.}~\bibnamefont {Olshanii}},\ }\href
  {\doibase 10.1038/nature06838} {\bibfield  {journal} {\bibinfo  {journal}
  {Nature}\ }\textbf {\bibinfo {volume} {452}},\ \bibinfo {pages} {854}
  (\bibinfo {year} {2008})}\BibitemShut {NoStop}%
\bibitem [{\citenamefont {Deutsch}(1991)}]{DeutschETH}%
  \BibitemOpen
  \bibfield  {author} {\bibinfo {author} {\bibfnamefont {J.~M.}\ \bibnamefont
  {Deutsch}},\ }\href@noop {} {\bibfield  {journal} {\bibinfo  {journal} {Phys.
  Rev. A}\ }\textbf {\bibinfo {volume} {43}},\ \bibinfo {pages} {2146}
  (\bibinfo {year} {1991})}\BibitemShut {NoStop}%
\bibitem [{\citenamefont {Srednicki}(1994)}]{SrednickiETH}%
  \BibitemOpen
  \bibfield  {author} {\bibinfo {author} {\bibfnamefont {M.}~\bibnamefont
  {Srednicki}},\ }\href@noop {} {\bibfield  {journal} {\bibinfo  {journal}
  {Phys. Rev. E}\ }\textbf {\bibinfo {volume} {50}},\ \bibinfo {pages} {888}
  (\bibinfo {year} {1994})}\BibitemShut {NoStop}%
\bibitem [{\citenamefont {Sutherland}(2004)}]{Sutherland}%
  \BibitemOpen
  \bibfield  {author} {\bibinfo {author} {\bibfnamefont {B.}~\bibnamefont
  {Sutherland}},\ }\href {http://books.google.com/books?id=aVUdnwEACAAJ} {\emph
  {\bibinfo {title} {Beautiful Models: 70 Years of Exactly Solved Quantum
  Many-body Problems}}}\ (\bibinfo  {publisher} {World Scientific},\ \bibinfo
  {year} {2004})\BibitemShut {NoStop}%
\bibitem [{\citenamefont {Kinoshita}\ \emph {et~al.}(2006)\citenamefont
  {Kinoshita}, \citenamefont {Wenger},\ and\ \citenamefont
  {Weiss}}]{Kinoshita06}%
  \BibitemOpen
  \bibfield  {author} {\bibinfo {author} {\bibfnamefont {T.}~\bibnamefont
  {Kinoshita}}, \bibinfo {author} {\bibfnamefont {T.}~\bibnamefont {Wenger}}, \
  and\ \bibinfo {author} {\bibfnamefont {D.~S.}\ \bibnamefont {Weiss}},\ }\href
  {\doibase 10.1038/nature04693} {\bibfield  {journal} {\bibinfo  {journal}
  {Nature}\ }\textbf {\bibinfo {volume} {440}},\ \bibinfo {pages} {900}
  (\bibinfo {year} {2006})}\BibitemShut {NoStop}%
\bibitem [{\citenamefont {Rigol}\ \emph {et~al.}(2007)\citenamefont {Rigol},
  \citenamefont {Dunjko}, \citenamefont {Yurovsky},\ and\ \citenamefont
  {Olshanii}}]{Rigol07}%
  \BibitemOpen
  \bibfield  {author} {\bibinfo {author} {\bibfnamefont {M.}~\bibnamefont
  {Rigol}}, \bibinfo {author} {\bibfnamefont {V.}~\bibnamefont {Dunjko}},
  \bibinfo {author} {\bibfnamefont {V.}~\bibnamefont {Yurovsky}}, \ and\
  \bibinfo {author} {\bibfnamefont {M.}~\bibnamefont {Olshanii}},\ }\href
  {\doibase 10.1103/PhysRevLett.98.050405} {\bibfield  {journal} {\bibinfo
  {journal} {Phys. Rev. Lett.}\ }\textbf {\bibinfo {volume} {98}},\ \bibinfo
  {pages} {050405} (\bibinfo {year} {2007})}\BibitemShut {NoStop}%
\bibitem [{\citenamefont {Calabrese}\ and\ \citenamefont {Cardy}(2007)}]{GGE2}%
  \BibitemOpen
  \bibfield  {author} {\bibinfo {author} {\bibfnamefont {P.}~\bibnamefont
  {Calabrese}}\ and\ \bibinfo {author} {\bibfnamefont {J.}~\bibnamefont
  {Cardy}},\ }\href {http://stacks.iop.org/1742-5468/2007/i=06/a=P06008}
  {\bibfield  {journal} {\bibinfo  {journal} {Journal of Statistical Mechanics:
  Theory and Experiment}\ }\textbf {\bibinfo {volume} {2007}},\ \bibinfo
  {pages} {P06008} (\bibinfo {year} {2007})}\BibitemShut {NoStop}%
\bibitem [{\citenamefont {Iucci}\ and\ \citenamefont {Cazalilla}(2009)}]{GGE3}%
  \BibitemOpen
  \bibfield  {author} {\bibinfo {author} {\bibfnamefont {A.}~\bibnamefont
  {Iucci}}\ and\ \bibinfo {author} {\bibfnamefont {M.~A.}\ \bibnamefont
  {Cazalilla}},\ }\href {\doibase 10.1103/PhysRevA.80.063619} {\bibfield
  {journal} {\bibinfo  {journal} {Phys. Rev. A}\ }\textbf {\bibinfo {volume}
  {80}},\ \bibinfo {pages} {063619} (\bibinfo {year} {2009})}\BibitemShut
  {NoStop}%
\bibitem [{\citenamefont {Basko}\ \emph {et~al.}(2006)\citenamefont {Basko},
  \citenamefont {Aleiner},\ and\ \citenamefont {Altshuler}}]{Basko06}%
  \BibitemOpen
  \bibfield  {author} {\bibinfo {author} {\bibfnamefont {D.}~\bibnamefont
  {Basko}}, \bibinfo {author} {\bibfnamefont {I.}~\bibnamefont {Aleiner}}, \
  and\ \bibinfo {author} {\bibfnamefont {B.}~\bibnamefont {Altshuler}},\ }\href
  {\doibase 10.1016/j.aop.2005.11.014} {\bibfield  {journal} {\bibinfo
  {journal} {Annals of Physics}\ }\textbf {\bibinfo {volume} {321}},\ \bibinfo
  {pages} {1126 } (\bibinfo {year} {2006})}\BibitemShut {NoStop}%
\bibitem [{\citenamefont {Gornyi}\ \emph {et~al.}(2005)\citenamefont {Gornyi},
  \citenamefont {Mirlin},\ and\ \citenamefont {Polyakov}}]{Mirlin05}%
  \BibitemOpen
  \bibfield  {author} {\bibinfo {author} {\bibfnamefont {I.~V.}\ \bibnamefont
  {Gornyi}}, \bibinfo {author} {\bibfnamefont {A.~D.}\ \bibnamefont {Mirlin}},
  \ and\ \bibinfo {author} {\bibfnamefont {D.~G.}\ \bibnamefont {Polyakov}},\
  }\href {\doibase 10.1103/PhysRevLett.95.206603} {\bibfield  {journal}
  {\bibinfo  {journal} {Phys. Rev. Lett.}\ }\textbf {\bibinfo {volume} {95}},\
  \bibinfo {pages} {206603} (\bibinfo {year} {2005})}\BibitemShut {NoStop}%
\bibitem [{\citenamefont {Oganesyan}\ and\ \citenamefont
  {Huse}(2007)}]{OganesyanHuse}%
  \BibitemOpen
  \bibfield  {author} {\bibinfo {author} {\bibfnamefont {V.}~\bibnamefont
  {Oganesyan}}\ and\ \bibinfo {author} {\bibfnamefont {D.~A.}\ \bibnamefont
  {Huse}},\ }\href {\doibase 10.1103/PhysRevB.75.155111} {\bibfield  {journal}
  {\bibinfo  {journal} {Phys. Rev. B}\ }\textbf {\bibinfo {volume} {75}},\
  \bibinfo {pages} {155111} (\bibinfo {year} {2007})}\BibitemShut {NoStop}%
\bibitem [{\citenamefont {Pal}\ and\ \citenamefont {Huse}(2010)}]{PalHuse}%
  \BibitemOpen
  \bibfield  {author} {\bibinfo {author} {\bibfnamefont {A.}~\bibnamefont
  {Pal}}\ and\ \bibinfo {author} {\bibfnamefont {D.~A.}\ \bibnamefont {Huse}},\
  }\href {\doibase 10.1103/PhysRevB.82.174411} {\bibfield  {journal} {\bibinfo
  {journal} {Phys. Rev. B}\ }\textbf {\bibinfo {volume} {82}},\ \bibinfo
  {pages} {174411} (\bibinfo {year} {2010})}\BibitemShut {NoStop}%
\bibitem [{\citenamefont {Anderson}(1958)}]{Anderson58}%
  \BibitemOpen
  \bibfield  {author} {\bibinfo {author} {\bibfnamefont {P.~W.}\ \bibnamefont
  {Anderson}},\ }\href {\doibase 10.1103/PhysRev.109.1492} {\bibfield
  {journal} {\bibinfo  {journal} {Phys. Rev.}\ }\textbf {\bibinfo {volume}
  {109}},\ \bibinfo {pages} {1492} (\bibinfo {year} {1958})}\BibitemShut
  {NoStop}%
\bibitem [{\citenamefont {Serbyn}\ \emph
  {et~al.}(2013{\natexlab{a}})\citenamefont {Serbyn}, \citenamefont
  {Papi\ifmmode~\acute{c}\else \'{c}\fi{}},\ and\ \citenamefont
  {Abanin}}]{Serbyn13-1}%
  \BibitemOpen
  \bibfield  {author} {\bibinfo {author} {\bibfnamefont {M.}~\bibnamefont
  {Serbyn}}, \bibinfo {author} {\bibfnamefont {Z.}~\bibnamefont
  {Papi\ifmmode~\acute{c}\else \'{c}\fi{}}}, \ and\ \bibinfo {author}
  {\bibfnamefont {D.~A.}\ \bibnamefont {Abanin}},\ }\href {\doibase
  10.1103/PhysRevLett.111.127201} {\bibfield  {journal} {\bibinfo  {journal}
  {Phys. Rev. Lett.}\ }\textbf {\bibinfo {volume} {111}},\ \bibinfo {pages}
  {127201} (\bibinfo {year} {2013}{\natexlab{a}})}\BibitemShut {NoStop}%
\bibitem [{\citenamefont {Huse}\ and\ \citenamefont
  {Oganesyan}(2013)}]{HuseOganesyan13}%
  \BibitemOpen
  \bibfield  {author} {\bibinfo {author} {\bibfnamefont {D.~A.}\ \bibnamefont
  {Huse}}\ and\ \bibinfo {author} {\bibfnamefont {V.}~\bibnamefont
  {Oganesyan}},\ }\href@noop {} {\bibfield  {journal} {\bibinfo  {journal}
  {ArXiv e-prints}\ } (\bibinfo {year} {2013})},\ \Eprint
  {http://arxiv.org/abs/arXiv:1305.4915} {arXiv:1305.4915} \BibitemShut
  {NoStop}%
\bibitem [{Note1()}]{Note1}%
  \BibitemOpen
  \bibinfo {note} {The notion of locality used here is different from the one
  used in conventional 1D integrable systems. Here, locality implies that an
  operator acts non-trivially on $\sim O(1)$ degrees of freedom. In this sense,
  a spin operator on a given site is local, whereas the Hamiltonian is not.
  Integrals of motion in the context of integrable systems are called local
  when they can be expressed as a sum of a local operators, and both examples
  above satisfy this definition.}\BibitemShut {Stop}%
\bibitem [{\citenamefont {Serbyn}\ \emph
  {et~al.}(2013{\natexlab{b}})\citenamefont {Serbyn}, \citenamefont
  {Papi\ifmmode~\acute{c}\else \'{c}\fi{}},\ and\ \citenamefont
  {Abanin}}]{Serbyn13-2}%
  \BibitemOpen
  \bibfield  {author} {\bibinfo {author} {\bibfnamefont {M.}~\bibnamefont
  {Serbyn}}, \bibinfo {author} {\bibfnamefont {Z.}~\bibnamefont
  {Papi\ifmmode~\acute{c}\else \'{c}\fi{}}}, \ and\ \bibinfo {author}
  {\bibfnamefont {D.~A.}\ \bibnamefont {Abanin}},\ }\href {\doibase
  10.1103/PhysRevLett.110.260601} {\bibfield  {journal} {\bibinfo  {journal}
  {Phys. Rev. Lett.}\ }\textbf {\bibinfo {volume} {110}},\ \bibinfo {pages}
  {260601} (\bibinfo {year} {2013}{\natexlab{b}})}\BibitemShut {NoStop}%
\bibitem [{\citenamefont {Vosk}\ and\ \citenamefont {Altman}(2013)}]{Altman13}%
  \BibitemOpen
  \bibfield  {author} {\bibinfo {author} {\bibfnamefont {R.}~\bibnamefont
  {Vosk}}\ and\ \bibinfo {author} {\bibfnamefont {E.}~\bibnamefont {Altman}},\
  }\href {\doibase 10.1103/PhysRevLett.110.067204} {\bibfield  {journal}
  {\bibinfo  {journal} {Phys. Rev. Lett.}\ }\textbf {\bibinfo {volume} {110}},\
  \bibinfo {pages} {067204} (\bibinfo {year} {2013})}\BibitemShut {NoStop}%
\bibitem [{\citenamefont {Znidaric}\ \emph {et~al.}(2008)\citenamefont
  {Znidaric}, \citenamefont {Prosen},\ and\ \citenamefont
  {Prelovsek}}]{Znidaric08}%
  \BibitemOpen
  \bibfield  {author} {\bibinfo {author} {\bibfnamefont {M.}~\bibnamefont
  {Znidaric}}, \bibinfo {author} {\bibfnamefont {T.}~\bibnamefont {Prosen}}, \
  and\ \bibinfo {author} {\bibfnamefont {P.}~\bibnamefont {Prelovsek}},\ }\href
  {\doibase 10.1103/PhysRevB.77.064426} {\bibfield  {journal} {\bibinfo
  {journal} {Phys. Rev. B}\ }\textbf {\bibinfo {volume} {77}},\ \bibinfo
  {pages} {064426} (\bibinfo {year} {2008})}\BibitemShut {NoStop}%
\bibitem [{\citenamefont {Bardarson}\ \emph {et~al.}(2012)\citenamefont
  {Bardarson}, \citenamefont {Pollmann},\ and\ \citenamefont
  {Moore}}]{Moore12}%
  \BibitemOpen
  \bibfield  {author} {\bibinfo {author} {\bibfnamefont {J.~H.}\ \bibnamefont
  {Bardarson}}, \bibinfo {author} {\bibfnamefont {F.}~\bibnamefont {Pollmann}},
  \ and\ \bibinfo {author} {\bibfnamefont {J.~E.}\ \bibnamefont {Moore}},\
  }\href {\doibase 10.1103/PhysRevLett.109.017202} {\bibfield  {journal}
  {\bibinfo  {journal} {Phys. Rev. Lett.}\ }\textbf {\bibinfo {volume} {109}},\
  \bibinfo {pages} {017202} (\bibinfo {year} {2012})}\BibitemShut {NoStop}%
\bibitem [{\citenamefont {Lieb}\ and\ \citenamefont
  {Robinson}(1972)}]{LiebRobinson}%
  \BibitemOpen
  \bibfield  {author} {\bibinfo {author} {\bibfnamefont {E.~H.}\ \bibnamefont
  {Lieb}}\ and\ \bibinfo {author} {\bibfnamefont {D.}~\bibnamefont
  {Robinson}},\ }\href@noop {} {\bibfield  {journal} {\bibinfo  {journal}
  {Commun. Math. Phys.}\ }\textbf {\bibinfo {volume} {28}},\ \bibinfo {pages}
  {251} (\bibinfo {year} {1972})}\BibitemShut {NoStop}%
\bibitem [{\citenamefont {Cheneau}\ \emph {et~al.}(2012)\citenamefont
  {Cheneau}, \citenamefont {Barmettler}, \citenamefont {Poletti}, \citenamefont
  {Endres}, \citenamefont {Schausz}, \citenamefont {Fukuhara}, \citenamefont
  {Gross}, \citenamefont {Bloch}, \citenamefont {Kollath},\ and\ \citenamefont
  {Kuhr}}]{Bloch12}%
  \BibitemOpen
  \bibfield  {author} {\bibinfo {author} {\bibfnamefont {M.}~\bibnamefont
  {Cheneau}}, \bibinfo {author} {\bibfnamefont {P.}~\bibnamefont {Barmettler}},
  \bibinfo {author} {\bibfnamefont {D.}~\bibnamefont {Poletti}}, \bibinfo
  {author} {\bibfnamefont {M.}~\bibnamefont {Endres}}, \bibinfo {author}
  {\bibfnamefont {P.}~\bibnamefont {Schausz}}, \bibinfo {author} {\bibfnamefont
  {T.}~\bibnamefont {Fukuhara}}, \bibinfo {author} {\bibfnamefont
  {C.}~\bibnamefont {Gross}}, \bibinfo {author} {\bibfnamefont
  {I.}~\bibnamefont {Bloch}}, \bibinfo {author} {\bibfnamefont
  {C.}~\bibnamefont {Kollath}}, \ and\ \bibinfo {author} {\bibfnamefont
  {S.}~\bibnamefont {Kuhr}},\ }\href {\doibase 10.1038/nature10748} {\bibfield
  {journal} {\bibinfo  {journal} {Nature}\ }\textbf {\bibinfo {volume} {481}},\
  \bibinfo {pages} {484} (\bibinfo {year} {2012})}\BibitemShut {NoStop}%
\bibitem [{\citenamefont {Kim}\ and\ \citenamefont {Huse}(2013)}]{KimHuse13}%
  \BibitemOpen
  \bibfield  {author} {\bibinfo {author} {\bibfnamefont {H.}~\bibnamefont
  {Kim}}\ and\ \bibinfo {author} {\bibfnamefont {D.~A.}\ \bibnamefont {Huse}},\
  }\href {\doibase 10.1103/PhysRevLett.111.127205} {\bibfield  {journal}
  {\bibinfo  {journal} {Phys. Rev. Lett.}\ }\textbf {\bibinfo {volume} {111}},\
  \bibinfo {pages} {127205} (\bibinfo {year} {2013})}\BibitemShut {NoStop}%
\bibitem [{\citenamefont {{Serbyn}}\ \emph {et~al.}(2014)\citenamefont
  {{Serbyn}}, \citenamefont {{Knap}}, \citenamefont {{Gopalakrishnan}},
  \citenamefont {{Papi{\'c}}}, \citenamefont {{Yao}}, \citenamefont
  {{Laumann}}, \citenamefont {{Abanin}}, \citenamefont {{Lukin}},\ and\
  \citenamefont {{Demler}}}]{Serbyn14}%
  \BibitemOpen
  \bibfield  {author} {\bibinfo {author} {\bibfnamefont {M.}~\bibnamefont
  {{Serbyn}}}, \bibinfo {author} {\bibfnamefont {M.}~\bibnamefont {{Knap}}},
  \bibinfo {author} {\bibfnamefont {S.}~\bibnamefont {{Gopalakrishnan}}},
  \bibinfo {author} {\bibfnamefont {Z.}~\bibnamefont {{Papi{\'c}}}}, \bibinfo
  {author} {\bibfnamefont {N.~Y.}\ \bibnamefont {{Yao}}}, \bibinfo {author}
  {\bibfnamefont {C.~R.}\ \bibnamefont {{Laumann}}}, \bibinfo {author}
  {\bibfnamefont {D.~A.}\ \bibnamefont {{Abanin}}}, \bibinfo {author}
  {\bibfnamefont {M.~D.}\ \bibnamefont {{Lukin}}}, \ and\ \bibinfo {author}
  {\bibfnamefont {E.~A.}\ \bibnamefont {{Demler}}},\ }\href@noop {} {\bibfield
  {journal} {\bibinfo  {journal} {ArXiv e-prints}\ } (\bibinfo {year}
  {2014})},\ \Eprint {http://arxiv.org/abs/1403.0693} {arXiv:1403.0693
  [cond-mat.dis-nn]} \BibitemShut {NoStop}%
\bibitem [{\citenamefont {Vasseur}\ \emph {et~al.}(2014)\citenamefont
  {Vasseur}, \citenamefont {Parameswaran},\ and\ \citenamefont
  {Moore}}]{Vasseur14}%
  \BibitemOpen
  \bibfield  {author} {\bibinfo {author} {\bibfnamefont {R.}~\bibnamefont
  {Vasseur}}, \bibinfo {author} {\bibfnamefont {S.~A.}\ \bibnamefont
  {Parameswaran}}, \ and\ \bibinfo {author} {\bibfnamefont {J.~E.}\
  \bibnamefont {Moore}},\ }\href@noop {} {\bibfield  {journal} {\bibinfo
  {journal} {ArXiv e-prints}\ } (\bibinfo {year} {2014})},\ \Eprint
  {http://arxiv.org/abs/arXiv:1407.4476} {arXiv:arXiv:1407.4476
  [cond-mat.dis-nn]} \BibitemShut {NoStop}%
\bibitem [{\citenamefont {Pasienski}\ \emph {et~al.}(2010)\citenamefont
  {Pasienski}, \citenamefont {McKay}, \citenamefont {White},\ and\
  \citenamefont {DeMarco}}]{DeMarco10}%
  \BibitemOpen
  \bibfield  {author} {\bibinfo {author} {\bibfnamefont {M.}~\bibnamefont
  {Pasienski}}, \bibinfo {author} {\bibfnamefont {D.}~\bibnamefont {McKay}},
  \bibinfo {author} {\bibfnamefont {M.}~\bibnamefont {White}}, \ and\ \bibinfo
  {author} {\bibfnamefont {B.}~\bibnamefont {DeMarco}},\ }\href@noop {}
  {\bibfield  {journal} {\bibinfo  {journal} {Nat. Phys.}\ }\textbf {\bibinfo
  {volume} {6}},\ \bibinfo {pages} {677} (\bibinfo {year} {2010})}\BibitemShut
  {NoStop}%
\bibitem [{\citenamefont {D'Errico}\ \emph {et~al.}(2013)\citenamefont
  {D'Errico}, \citenamefont {Moratti}, \citenamefont {Lucioni}, \citenamefont
  {Tanzi}, \citenamefont {Deissler},\ and\ \citenamefont
  {Inguscio}}]{Inguscio13}%
  \BibitemOpen
  \bibfield  {author} {\bibinfo {author} {\bibfnamefont {C.}~\bibnamefont
  {D'Errico}}, \bibinfo {author} {\bibfnamefont {M.}~\bibnamefont {Moratti}},
  \bibinfo {author} {\bibfnamefont {E.}~\bibnamefont {Lucioni}}, \bibinfo
  {author} {\bibfnamefont {L.}~\bibnamefont {Tanzi}}, \bibinfo {author}
  {\bibfnamefont {B.}~\bibnamefont {Deissler}}, \ and\ \bibinfo {author}
  {\bibfnamefont {M.}~\bibnamefont {Inguscio}},\ }\href@noop {} {\bibfield
  {journal} {\bibinfo  {journal} {New J. Phys.}\ }\textbf {\bibinfo {volume}
  {15}},\ \bibinfo {pages} {045007} (\bibinfo {year} {2013})}\BibitemShut
  {NoStop}%
\bibitem [{\citenamefont {Kondov}\ \emph {et~al.}(2013)\citenamefont {Kondov},
  \citenamefont {McGehee},\ and\ \citenamefont {DeMarco}}]{DeMarco13}%
  \BibitemOpen
  \bibfield  {author} {\bibinfo {author} {\bibfnamefont {S.~S.}\ \bibnamefont
  {Kondov}}, \bibinfo {author} {\bibfnamefont {W.~R.}\ \bibnamefont {McGehee}},
  \ and\ \bibinfo {author} {\bibfnamefont {B.}~\bibnamefont {DeMarco}},\
  }\href@noop {} {\bibfield  {journal} {\bibinfo  {journal} {Arxiv e-prints}\ }
  (\bibinfo {year} {2013})},\ \Eprint {http://arxiv.org/abs/1305.6072}
  {1305.6072} \BibitemShut {NoStop}%
\bibitem [{\citenamefont {Childress}\ \emph {et~al.}(2006)\citenamefont
  {Childress}, \citenamefont {Gurudev~Dutt}, \citenamefont {Taylor},
  \citenamefont {Zibrov}, \citenamefont {Jelezko}, \citenamefont {Wrachtrup},
  \citenamefont {Hemmer},\ and\ \citenamefont {Lukin}}]{Lukin06}%
  \BibitemOpen
  \bibfield  {author} {\bibinfo {author} {\bibfnamefont {L.}~\bibnamefont
  {Childress}}, \bibinfo {author} {\bibfnamefont {M.~V.}\ \bibnamefont
  {Gurudev~Dutt}}, \bibinfo {author} {\bibfnamefont {J.~M.}\ \bibnamefont
  {Taylor}}, \bibinfo {author} {\bibfnamefont {A.~S.}\ \bibnamefont {Zibrov}},
  \bibinfo {author} {\bibfnamefont {F.}~\bibnamefont {Jelezko}}, \bibinfo
  {author} {\bibfnamefont {J.}~\bibnamefont {Wrachtrup}}, \bibinfo {author}
  {\bibfnamefont {P.~R.}\ \bibnamefont {Hemmer}}, \ and\ \bibinfo {author}
  {\bibfnamefont {M.~D.}\ \bibnamefont {Lukin}},\ }\href {\doibase
  10.1126/science.1131871} {\bibfield  {journal} {\bibinfo  {journal}
  {Science}\ }\textbf {\bibinfo {volume} {314}},\ \bibinfo {pages} {281}
  (\bibinfo {year} {2006})}\BibitemShut {NoStop}%
\bibitem [{\citenamefont {Neumann}\ \emph {et~al.}(2008)\citenamefont
  {Neumann}, \citenamefont {Mizuochi}, \citenamefont {Rempp}, \citenamefont
  {Hemmer}, \citenamefont {Watanabe}, \citenamefont {Yamasami}, \citenamefont
  {Jacques}, \citenamefont {Gaebel}, \citenamefont {Jelezko},\ and\
  \citenamefont {Wrachtrup}}]{Wrachtrup08}%
  \BibitemOpen
  \bibfield  {author} {\bibinfo {author} {\bibfnamefont {P.}~\bibnamefont
  {Neumann}}, \bibinfo {author} {\bibfnamefont {N.}~\bibnamefont {Mizuochi}},
  \bibinfo {author} {\bibfnamefont {F.}~\bibnamefont {Rempp}}, \bibinfo
  {author} {\bibfnamefont {P.}~\bibnamefont {Hemmer}}, \bibinfo {author}
  {\bibfnamefont {H.}~\bibnamefont {Watanabe}}, \bibinfo {author}
  {\bibfnamefont {S.}~\bibnamefont {Yamasami}}, \bibinfo {author}
  {\bibfnamefont {V.}~\bibnamefont {Jacques}}, \bibinfo {author} {\bibfnamefont
  {T.}~\bibnamefont {Gaebel}}, \bibinfo {author} {\bibfnamefont
  {F.}~\bibnamefont {Jelezko}}, \ and\ \bibinfo {author} {\bibfnamefont
  {J.}~\bibnamefont {Wrachtrup}},\ }\href@noop {} {\bibfield  {journal}
  {\bibinfo  {journal} {Science}\ }\textbf {\bibinfo {volume} {320}},\ \bibinfo
  {pages} {1326} (\bibinfo {year} {2008})}\BibitemShut {NoStop}%
\bibitem [{\citenamefont {Bauer}\ and\ \citenamefont
  {Nayak}(2013)}]{BauerNayak}%
  \BibitemOpen
  \bibfield  {author} {\bibinfo {author} {\bibfnamefont {B.}~\bibnamefont
  {Bauer}}\ and\ \bibinfo {author} {\bibfnamefont {C.}~\bibnamefont {Nayak}},\
  }\href@noop {} {\bibfield  {journal} {\bibinfo  {journal} {J. Stat. Mech.}\
  ,\ \bibinfo {pages} {P09005}} (\bibinfo {year} {2013})}\BibitemShut {NoStop}%
\bibitem [{\citenamefont {Kj\"all}\ \emph {et~al.}(2014)\citenamefont
  {Kj\"all}, \citenamefont {Bardarson},\ and\ \citenamefont
  {Pollmann}}]{Kjall14}%
  \BibitemOpen
  \bibfield  {author} {\bibinfo {author} {\bibfnamefont {J.~A.}\ \bibnamefont
  {Kj\"all}}, \bibinfo {author} {\bibfnamefont {J.~H.}\ \bibnamefont
  {Bardarson}}, \ and\ \bibinfo {author} {\bibfnamefont {F.}~\bibnamefont
  {Pollmann}},\ }\href@noop {} {\bibfield  {journal} {\bibinfo  {journal}
  {ArXiv e-prints}\ } (\bibinfo {year} {2014})},\ \Eprint
  {http://arxiv.org/abs/arXiv:1403.1568} {arXiv:arXiv:1403.1568
  [cond-mat.dis-nn]} \BibitemShut {NoStop}%
\bibitem [{\citenamefont {Imbrie}(2014)}]{Imbrie14}%
  \BibitemOpen
  \bibfield  {author} {\bibinfo {author} {\bibfnamefont {J.~Z.}\ \bibnamefont
  {Imbrie}},\ }\href@noop {} {\bibfield  {journal} {\bibinfo  {journal} {Arxiv
  e-prints}\ } (\bibinfo {year} {2014})},\ \Eprint
  {http://arxiv.org/abs/1403.7837} {1403.7837} \BibitemShut {NoStop}%
\bibitem [{\citenamefont {Chandran}\ \emph {et~al.}(2014)\citenamefont
  {Chandran}, \citenamefont {Kim}, \citenamefont {Vidal},\ and\ \citenamefont
  {Abanin}}]{Chandran14}%
  \BibitemOpen
  \bibfield  {author} {\bibinfo {author} {\bibfnamefont {A.}~\bibnamefont
  {Chandran}}, \bibinfo {author} {\bibfnamefont {I.~H.}\ \bibnamefont {Kim}},
  \bibinfo {author} {\bibfnamefont {G.}~\bibnamefont {Vidal}}, \ and\ \bibinfo
  {author} {\bibfnamefont {D.~A.}\ \bibnamefont {Abanin}},\ }\href@noop {}
  {\bibfield  {journal} {\bibinfo  {journal} {ArXiv e-prints}\ } (\bibinfo
  {year} {2014})},\ \Eprint {http://arxiv.org/abs/arXiv:1407.8480}
  {arXiv:arXiv:1407.8480 [cond-mat.dis-nn]} \BibitemShut {NoStop}%
\bibitem [{\citenamefont {Ros}\ \emph {et~al.}(2014)\citenamefont {Ros},
  \citenamefont {Mueller},\ and\ \citenamefont {Scardicchio}}]{Ros14}%
  \BibitemOpen
  \bibfield  {author} {\bibinfo {author} {\bibfnamefont {V.}~\bibnamefont
  {Ros}}, \bibinfo {author} {\bibfnamefont {M.}~\bibnamefont {Mueller}}, \ and\
  \bibinfo {author} {\bibfnamefont {A.}~\bibnamefont {Scardicchio}},\
  }\href@noop {} {\bibfield  {journal} {\bibinfo  {journal} {ArXiv e-prints}\ }
  (\bibinfo {year} {2014})},\ \Eprint {http://arxiv.org/abs/arXiv:1406.2175}
  {arXiv:arXiv:1406.2175 [cond-mat.dis-nn]} \BibitemShut {NoStop}%
\bibitem [{\citenamefont {Hamza}\ \emph {et~al.}(2014)\citenamefont {Hamza},
  \citenamefont {Sims},\ and\ \citenamefont {Stolz}}]{Hamza}%
  \BibitemOpen
  \bibfield  {author} {\bibinfo {author} {\bibfnamefont {E.}~\bibnamefont
  {Hamza}}, \bibinfo {author} {\bibfnamefont {R.}~\bibnamefont {Sims}}, \ and\
  \bibinfo {author} {\bibfnamefont {G.}~\bibnamefont {Stolz}},\ }\href@noop {}
  {\bibfield  {journal} {\bibinfo  {journal} {ArXiv e-prints}\ } (\bibinfo
  {year} {2014})},\ \Eprint {http://arxiv.org/abs/arXiv:1108.3811}
  {arXiv:arXiv:1108.3811 [math-ph]} \BibitemShut {NoStop}%
\bibitem [{\citenamefont {{Kim \emph{et. al.}}}(2014)}]{Kim-prep}%
  \BibitemOpen
  \bibfield  {author} {\bibinfo {author} {\bibfnamefont {I.~H.}\ \bibnamefont
  {{Kim \emph{et. al.}}}},\ }\href@noop {} {} (\bibinfo {year} {2014}),\
  \bibinfo {note} {in preparation}\BibitemShut {NoStop}%
\bibitem [{\citenamefont {{Bahri}}\ \emph {et~al.}(2013)\citenamefont
  {{Bahri}}, \citenamefont {{Vosk}}, \citenamefont {{Altman}},\ and\
  \citenamefont {{Vishwanath}}}]{Bahri}%
  \BibitemOpen
  \bibfield  {author} {\bibinfo {author} {\bibfnamefont {Y.}~\bibnamefont
  {{Bahri}}}, \bibinfo {author} {\bibfnamefont {R.}~\bibnamefont {{Vosk}}},
  \bibinfo {author} {\bibfnamefont {E.}~\bibnamefont {{Altman}}}, \ and\
  \bibinfo {author} {\bibfnamefont {A.}~\bibnamefont {{Vishwanath}}},\
  }\href@noop {} {\bibfield  {journal} {\bibinfo  {journal} {ArXiv e-prints}\ }
  (\bibinfo {year} {2013})},\ \Eprint {http://arxiv.org/abs/1307.4092}
  {arXiv:1307.4092 [cond-mat.dis-nn]} \BibitemShut {NoStop}%
\bibitem [{\citenamefont {Nandkishore}\ \emph {et~al.}(2014)\citenamefont
  {Nandkishore}, \citenamefont {Gopalakrishnan},\ and\ \citenamefont
  {Huse}}]{Nandkishore14}%
  \BibitemOpen
  \bibfield  {author} {\bibinfo {author} {\bibfnamefont {R.}~\bibnamefont
  {Nandkishore}}, \bibinfo {author} {\bibfnamefont {S.}~\bibnamefont
  {Gopalakrishnan}}, \ and\ \bibinfo {author} {\bibfnamefont {D.~A.}\
  \bibnamefont {Huse}},\ }\href@noop {} {\bibfield  {journal} {\bibinfo
  {journal} {ArXiv e-prints}\ } (\bibinfo {year} {2014})},\ \Eprint
  {http://arxiv.org/abs/arXiv:1402.5971} {arXiv:arXiv:1402.5971
  [cond-mat.stat-mech]} \BibitemShut {NoStop}%
\bibitem [{\citenamefont {Johri}\ \emph {et~al.}(2014)\citenamefont {Johri},
  \citenamefont {Nandkishore},\ and\ \citenamefont {Bhatt}}]{Johri14}%
  \BibitemOpen
  \bibfield  {author} {\bibinfo {author} {\bibfnamefont {S.}~\bibnamefont
  {Johri}}, \bibinfo {author} {\bibfnamefont {R.}~\bibnamefont {Nandkishore}},
  \ and\ \bibinfo {author} {\bibfnamefont {R.~N.}\ \bibnamefont {Bhatt}},\
  }\href@noop {} {\bibfield  {journal} {\bibinfo  {journal} {ArXiv e-prints}\ }
  (\bibinfo {year} {2014})},\ \Eprint {http://arxiv.org/abs/arXiv:1402.5971}
  {arXiv:arXiv:1402.5971 [cond-mat.dis-nn]} \BibitemShut {NoStop}%
\end{thebibliography}
\end{document}